\documentclass[english,aps,prl,reprint,superscriptaddress,floatfix,amsfonts]{revtex4-1}
\usepackage[latin9]{inputenc}
\usepackage{color}
\usepackage{bm}
\usepackage{multirow}
\usepackage{amsmath}
\usepackage{graphicx}

\makeatletter

\providecommand{\tabularnewline}{\\}

\usepackage{epstopdf}
\usepackage{dcolumn}
\usepackage{bm}
\usepackage{babel}
\usepackage{multirow}

\makeatother

\usepackage{babel}
\begin{document}
\title{Anisotropic superconductivity mediated by ferroelectric fluctuations
in cubic systems with spin-orbit coupling}
\date{\today}
\author{Maria N. Gastiasoro}
\altaffiliation[Present address: ]{ISC-CNR and Department of Physics, Sapienza University of Rome, Piazzale Aldo Moro 2, I-00185, Rome, Italy.}
\affiliation{School of Physics and Astronomy, University of Minnesota, Minneapolis
55455, USA}
\author{Tha\'{i}s V. Trevisan}
\altaffiliation[Present address: ]{Ames Laboratory, Ames, Iowa 50011, USA}
\affiliation{Instituto de F\'{i}sica Gleb Wataghin, Unicamp, Rua S\'ergio Buarque de
Holanda, 777, CEP 13083-859 Campinas, SP, Brazil}
\author{Rafael M. Fernandes}
\affiliation{School of Physics and Astronomy, University of Minnesota, Minneapolis
55455, USA}
\begin{abstract}
Motivated by the experimental observation that superconductivity in
bulk doped SrTiO$_{3}$ is enhanced as a putative ferroelectric quantum
critical point (FE-QCP) is approached, we study the pairing instability
of a cubic system in which electrons exchange low-energy ferroelectric
fluctuations. Instead of the gradient coupling to the lattice distortion
associated with ferroelectricity, we consider a direct coupling between
the electrons and the bosonic ferroelectric field that appears in
the presence of spin-orbit coupling. Working in the weak-coupling
regime, we find that the pairing interaction is dominated by the
soft transverse optical (TO) mode, resulting in a $T_{c}$ enhancement
upon approaching the FE-QCP. Focusing on even-parity states, we find
that although the $s$-wave state always wins, states with higher
Cooper-pair angular momentum become close competitors as the TO mode
softens. We show that the cubic anisotropy of the FE fluctuations
mixes the $s$-wave and $g$-wave states, resulting in a characteristic
anisotropy of the gap function. The gap anisotropy behaves non-monotonically
as the FE-QCP is approached: upon decreasing the TO mode frequency,
the gap anisotropy first changes sign and then increases in magnitude.
We discuss the possible applications of our results to the superconducting
state of SrTiO$_{3}$.
\end{abstract}
\maketitle

\section{Introduction}

The possibility that bosonic excitations other than phonons can promote
superconductivity has a long history. The fact that the pairing state
of high-$T_{c}$ cuprates is observed in close proximity to an antiferromagnetic
state has encouraged intense theoretical investigations about the
nature of the superconducting state mediated by antiferromagnetic
fluctuations~\citep{Pines91,Abanov2003,Scalapino12,Taillefer2010,Sachdev10,Wang17}. 
Beyond cuprates, the phase diagrams of certain heavy
fermions and iron pnictides have motivated theoretical studies of
pairing promoted by the exchange of ferromagnetic fluctuations \citep{Roussev01} and
nematic fluctuations \cite{Metlitski15,Lederer17}. 
More recently, the idea that ferroelectric fluctuations
can also mediate the formation of Cooper pairs~\citep{Edge15,Chandra17} has spurred considerable
interest, particularly in the context of doped bulk SrTiO$_{3}$ (STO) -- for a recent review, see Ref. \citep{review2019}.
Indeed, undoped STO is a semiconducting quantum paraelectric, i.e.
a material whose quantum fluctuations prevent the onset of the classical
ferroelectric ground state~\citep{Mueller79,Rowley2014}. Upon doping via oxygen vacancies or niobium
substitution, a superconducting dome emerges already at very small
carrier concentrations~\citep{Koonce67,Lin14critical,Bretz2019}. By tuning doped (i.e. metallic) STO towards
a ferroelectric transition, which can be accomplished via isotope
oxygen substitution~\citep{Stucky2016}, chemical substitution on the cation site~\citep{Rischau2017,Tomioka2019}, hydrostatic pressure~\citep{Rowley2018} or strain~\citep{Herrera2019,Russell2019}, it
is generally observed that the superconducting transition temperature
$T_{c}$ increases as the putative zero-temperature ferroelectric
transition is approached. This is particularly unexpected in the case
of $^{18}$O substituted STO~\citep{Stucky2016,Tomioka2019}, since the standard isotope effect would
predict a lower $T_{c}$ due to the fact that $^{18}$O is heavier
than $^{16}$O.

It is important to emphasize that a metal cannot sustain macroscopic
ferroelectricity due to screening effects. But in STO, the displacive
ferroelectric transition is accompanied by a structural transition
in which the crystal changes from centrosymmetric to non-centrosymmetric.
As a result, the ferroelectric transition is signaled not only by
a diverging dielectric constant~\citep{Weaver1959,Mueller79,Rowley2014}, but also by an accompanying softening
of a tranverse optical (TO) phonon mode~\citep{Cochran1960,Cowley64,Yamada69}. Strain fluctuations associated
with the TO mode persist even in the metallic phase, and thus can
mediate electron-electron interactions even if there are no macroscopic
dipole moments. Because locally the electric polarization is proportional
to the lattice displacement, we will refer to the pairing mechanism
as promoted by ferroelectric fluctuations, and to the ordered state
as a metallic ferroelectric state~\citep{Scott1974,Shi2013}.

Different theoretical models have been proposed to study the scenario
in which pairing is due to the exchange of such ferroelectric fluctuations.
Edge \emph{et al.} considered an effective model in which the zero-temperature
(i.e. quantum) ferroelectric phase transition in STO is described
in terms of a transverse field Ising model, which couples directly
via a Yukawa-type coupling to the electronic density~\citep{Edge15}. They obtained
a superconducting dome as the carrier concentration increases due
to the competition between the enhancement of the density of states
and the suppression of the soft TO mode upon doping. They also predicted
the aforementioned unusual isotope effect in $^{18}$O substituted
STO. One problem however is how the TO mode microscopically couples
to the electronic degrees of freedom. Because the TO mode is polar,
the standard electron-phonon matrix element gives a gradient coupling
between the lattice displacement and the electronic density. As a
result, the soft TO mode effectively decouples from the electronic
states, and the main contribution to the pairing interaction comes
from the associated longitudinal optical (LO) mode, which remains
massive even at the ferroelectric transition. W\"olfle and Balatsky
pointed out that the cubic anisotropy of the lattice allows for an
effective coupling between the TO mode and the electronic density
away from high-symmetry directions~\citep{Woelfle18}. However, such a coupling was later
argued to be too small~\citep{Ruhman19}.

An alternative coupling between the odd-parity TO mode and the electrons
occurs when spin-orbit coupling (SOC) is present~\citep{Fu15}. This allows for
a direct coupling between the fermions and the bosonic fields that
avoids the gradient coupling discussed above. Kozii and Fu recently
studied how superconductivity emerges in the general case of electrons
spin-orbit-coupled to a generic bosonic mode that breaks inversion
symmetry~\citep{KoziiFu15}. They found the interesting possibility of closely competing
even-parity and odd-parity pairing states. Kanasugi and Yanase considered
the same coupling to study the interplay between superconductivity
and long-range ferroelectric order in STO~\citep{Kanasugi18,Kanasugi19}. However, in their model,
the pairing interaction did not arise directly from the ferroelectric
fluctuations, i.e. long-range superconducting and ferroelectric orders
were treated as separate states.

Motivated by these previous investigations, in this paper we solve
the weak-coupling problem in which pairing is mediated by the exchange
of an inversion-symmetry-breaking vectorial bosonic field that couples
to the electrons via the SOC, as relevant for bulk STO. We explicitly
take into account the role of the cubic crystal-field anisotropy present in these
systems, an effect that has been largely unexplored in previous works.
We find that, in the singlet channel, the $s$-wave pairing channel
dominates, and the pairing interaction is strongly enhanced as the
ferroelectric transition is approached. Moreover, the subleading larger
angular momentum pseudospin-singlet channels ($d$-wave and $g$-wave) become
gradually more competitive as the TO mode becomes softer and the putative
ferroelectric quantum critical point (FE-QCP) is approached. Interestingly,
the lattice cubic anisotropy leads to an anisotropy of the gap function.
Although the anisotropy that we find is never large enough to induce
accidental nodes, it increases in magnitude and changes its sign as
the frequency of the TO mode decreases. As a result, the gap maxima
and gap minima switch locations around the Fermi surface as the FE-QCP
is approached.

We emphasize that the pairing solution studied here is in the weak-coupling
(i.e. BCS-like) limit, since the dynamics of the bosons are not taken
into account. This approximation is of course not valid close enough
to the FE-QCP, where the feedback effect of the fermions on the boson
dynamics is expected to play a crucial role. Moreover, in our approach,
because the main contribution to the pairing interaction comes from
the soft TO mode, whose energy is smaller than the Fermi energy, the
impact of the LO mode, whose energy is larger than the Fermi energy,
is negligible. Of course, the LO mode on its own can mediate pairing,
as discussed elsewhere \citep{Gastiasoro19}.
Despite these approximations, our work reveals
a clear qualitative evolution of the gap function as the FE-QCP is
approached. We thus discuss the possible implications of our results
to the elucidation of superconductivity in STO, highlighting the issues
that remain to be addressed to confirm whether this scenario is suitable.

The paper is organized as follows: in Sec.~\ref{sec:model} we introduce the model, which includes a spin-orbit mediated direct coupling between the low-energy fermions and the FE fluctuations, and obtain the effective pairing interaction and superconducting gap equation. We first solve for $T_c$ assuming rotational invariance in Sec.~\ref{sec:isotropic}, and investigate the close competition among even-parity channels as the TO mode softens. In Sec.~\ref{sec:cubic} we include the finite cubic crystal-field in the FE fluctuations and study the resulting anisotropy of the gap function. We show a characteristic sign change and subsequent growth of this anisotropy as the FE instability is approached. We summarize our results and discuss the possible applications to superconductivity in STO in Sec.~\ref{sec:discussion}. Appendices A and B expand on technical details of the main calculations.

\section{Low-energy model} \label{sec:model}

Our model consists of non-interacting fermions $c_{\mathbf{k}\alpha}$,
with momentum $\mathbf{k}$ and (pseudo-)spin projection $\alpha$,
coupled to ferroelectric (FE) excitations in a cubic lattice described
by the vector bosonic field $\boldsymbol{\phi}\left(\mathbf{q}\right)$.
The latter is parity-odd and time-reversal-even; as a result, it is
proportional to the polar lattice displacement $\mathbf{u}$ that
promotes a local polarization $\mathbf{P}\propto\mathbf{u}$ inside
the cubic unit cell. Since we are working in the metallic regime,
the system does not sustain a macroscopic polarization. Yet, as we
discussed above, we refer to $\boldsymbol{\phi}\left(\mathbf{q}\right)$
as the FE order parameter or polarization. Similarly, we refer to
the FE mode and the TO phonon mode interchangeably.

\subsection{Bosonic propagator: optical phonons}

The bosonic propagator describing the FE fluctuations in the disordered
state is given by~\citep{Roussev,Conduit,Woelfle18,Ruhman19,review2019}
\begin{equation}
\chi_{ij}^{-1}(\mathbf{q},i\Omega_{n})=E_{T}^{-1}\left[K_{i}(\mathbf{q})\delta_{ij}+M(q)\hat{q}_{i}\hat{q}_{j}\right]\text{ ,}\label{eq:D0}
\end{equation}

\noindent with Matsubara bosonic frequency $\Omega_{n}=2n\pi T$ ($n\in\mathbb{Z}$)
and: 
\begin{align}
 & K_{i}(\mathbf{q})=\Omega_{n}^{2}+\omega_{T}^{2}+E_{T}^{2}q^{2}+\varepsilon_{c}^{2}q_{i}^{2}\text{ ,}\\
 & M(q)=\omega_{L}^{2}-\omega_{T}^{2}-\left(E_{T}^{2}-E_{L}^{2}\right)q^{2}\text{ .}\label{eq:M}
\end{align}

Here, the Latin indices $i,j$ refer to components of the FE order
parameter $\phi_{i}$. The eigenvalues of Eq.(\ref{eq:D0}) give the
phonon dispersions of two transverse optical (TO) modes and one longitudinal
optical (LO) mode, which have the characteristic energy scales $E_{T}\equiv c_{T}\pi/a$
and $E_{L}\equiv c_{L}\pi/a$, respectively, with $a$ denoting the
cubic lattice constant. The quantities $c_{T}$ ($c_{L}$) and $\omega_{T}$
($\omega_{L}$) denote the transverse (longitudinal) phonon velocity
and the transverse (longitudinal) optical gap at the center of the
Brillouin zone, respectively. Importantly, $\omega_{L}^{2}=\omega_{T}^{2}+\omega_{p}^{2}$,
where $\omega_{p}$ is the ionic plasma frequency; this term arises
because the LO mode generates Coulomb energy. Thus, $\omega_{L}$
remains finite even when $\omega_{T}\rightarrow0$ at the FE transition.
Note that our choice of units is such that the propagator has dimensions
of inverse of energy, and the transferred momentum $\mathbf{q}$ has
units of $\pi/a$ in Eqs.(\ref{eq:D0})-(\ref{eq:M}).

\begin{figure}
\centering \includegraphics[width=\linewidth]{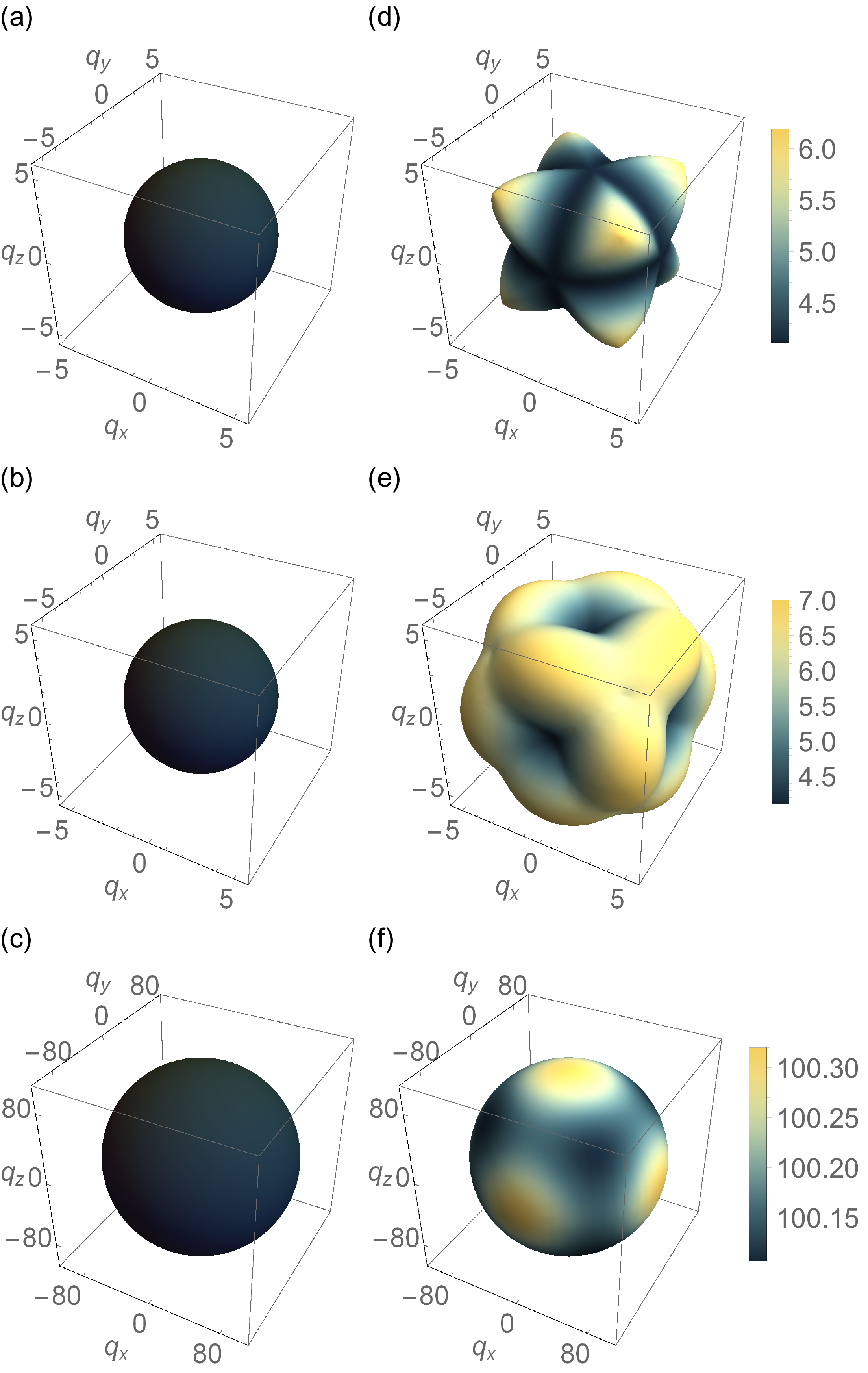}
\caption{Dispersions of the phonon propagator Eq. (1) at finite $q=0.1$ without
a crystal field $\varepsilon_{c}=0$ in panels (a)-(c) and with a finite
cubic crystal field $\varepsilon_{c}=2E_{T}$ in panels (d)-(f). (a)
and (b) are the purely transverse modes $\varpi_{T}(q)$ and (c) is
the purely longitudinal mode $\varpi_{L}(q)$. The eigenmodes of the
dispersions (d)-(f) are anisotropic; their polarizations are neither
purely longitudinal nor purely transverse except at high-symmetry
directions of the cubic crystal. In all panels $\omega_{T}=1$ meV,
$\omega_{L}=100$ meV, $E_{T}=40$ meV, and $E_{L}=E_{T}/10$.}
\label{fig:vals} 
\end{figure}

A crucial parameter in Eq. (\ref{eq:M}) is the cubic anisotropy term
$\varepsilon_{c}$, which arises from the crystal field effects of
the cubic lattice. If it is absent ($\varepsilon_{c}=0$), the phonon
propagator is rotationally invariant and can be rewritten in the form:

\begin{align}
\frac{\chi_{ij}^{-1}(\mathbf{q},i\Omega_{n})}{E_{T}^{-1}} & =\Omega_{n}^{2}\delta_{ij}+\left(\omega_{L}^{2}+E_{L}^{2}q^{2}\right)\hat{q}_{i}\hat{q}_{j}\label{eq:chi_iso}\\
 & +\left(\omega_{T}^{2}+E_{T}^{2}q^{2}\right)\left(\delta_{ij}-\hat{q}_{i}\hat{q}_{j}\right)\nonumber 
\end{align}
As a result, the eigenvalues split into a doubly-degenerate purely
transverse mode, $\varpi_{T}^{2}(\mathbf{q})=\omega_{T}^{2}+E_{T}^{2}q^{2}$,
and a purely longitudinal mode, $\varpi_{L}^{2}(\mathbf{q})=\omega_{L}^{2}+E_{L}^{2}q^{2}$.
A non-zero $\varepsilon_{c}$, on the other hand, breaks rotational
symmetry and mixes the longitudinal and transverse polarization of
the modes, except along high-symmetry directions~\citep{Woelfle18,Ruhman19}.
The dispersions themselves become anisotropic, as shown in Fig. \ref{fig:vals},
which contrasts the three isotropic modes for $\varepsilon_{c}=0$
(panels (a) to (c)) to the three anisotropic modes for $\varepsilon_{c}\neq0$
(panels (d) to (f)). As we explain later, the cubic anisotropy has
an important impact on the gap function in the superconducting state.
In this figure and in the remainder of the
text, we set the following parameters:
$\omega_{L}=100$ meV, $E_{T}=40$ meV, and $E_{L}=E_{T}/10$, which
fit well the neutron scattering data of Ref.~\cite{Yamada69}.

\subsection{Coupling to electronic degrees of freedom}

We now consider how the FE fluctuations discussed above couple to
the electronic degrees of freedom. For simplicity, we will restrict
our analysis to a single-band with dispersion $\xi_{\mathbf{k}}=k^{2}/2m-\mu$,
where $\mu$ is the chemical potential, as appropriate for dilute
STO. The most straightforward coupling between the FE bosonic field
$\boldsymbol{\phi}$ and the fermions $c_{\mathbf{k}\alpha}$ is via
the standard electron-phonon coupling. Due to the dipolar nature of
the phonons, this translates into a gradient coupling~\citep{Woelfle18,ArceGamboa18,Kedem18}.
As a result, there is no direct coupling between the modes with transverse
polarization and the electronic density. In the case where the cubic
anisotropy term vanishes, $\varepsilon_{c}=0$, this would imply a
complete decoupling from the soft TO mode. The presence of $\varepsilon_{c}\neq0$,
however, mixes transverse and longitudinal polarizations (except along
high-symmetry directions), allowing for an indirect coupling to the
TO mode \citep{Woelfle18}. Such a coupling, however, is expected
to be very small, particularly in the dilute regime of STO \citep{Ruhman19}.

In this work, we consider instead another allowed coupling, which
is present in systems with spin-orbit coupling (SOC), as previously
discussed in Refs. \citep{Fu15,KoziiFu15,Martin17,Kozii19}. The Hamiltonian
in this case is is given by

\begin{equation}
\hat{H}=\sum\limits _{\mathbf{k},\alpha}\xi_{\mathbf{k}}c_{\mathbf{k}\alpha}^{\dag}c_{\mathbf{k}\alpha}^{\null}+g\sum\limits _{\mathbf{q}}\sum\limits _{i}\phi_{i}(\mathbf{q})\hat{Q}_{i}(\mathbf{q})\text{.}\label{eq:H}
\end{equation}

\noindent where $g$ is a coupling constant with dimensions of energy.
Following the notation of Refs. \citep{KoziiFu15,Martin17}, the bilinear
electronic operator that couples directly to the parity-odd bosonic
field can be written as:

\noindent 
\begin{equation}
\hat{Q}_{i}(\mathbf{q})\equiv\sum\limits _{\mathbf{k},\alpha\beta}F_{i,\alpha\beta}(\mathbf{k},\mathbf{q})c_{\mathbf{k}+\mathbf{q},\alpha}^{\dag}c_{\mathbf{k}\beta}^{\null}\text{ ,}
\end{equation}
where the form factor is:

\noindent 
\begin{align}
 & F_{i,\alpha\beta}(\mathbf{k},\mathbf{q})=\frac{1}{2}\left[\Gamma_{i,\alpha\beta}(\mathbf{k}+\mathbf{q})+\Gamma_{i,\alpha\beta}(\mathbf{k})\right]\text{ ,}\label{eq:form_factor}\\
 & \Gamma_{i,\alpha\beta}(\mathbf{k})=[\hat{k}\times\bm{\sigma}_{\alpha\beta}]_{i}\text{ ,}\label{eq:gamma}
\end{align}

\noindent Recall that the Latin indices $i,j$ refer to the Cartesian
components of the bosonic fields, whereas the Greek indices $\alpha,\beta$
refer to the pseudospin components of the fermionic field. Moreover, $\sigma$
is a Pauli matrix. That such a term is allowed by symmetry follows
from the fact that $\hat{Q}_{i}(\mathbf{q})$ is even under time-reversal
but odd under inversion symmetry. The main question, of course, is
about the magnitude of the coupling constant $g$, which remains unsettled
in STO, to the best of our knowledge~\citep{Ruhman16}. Qualitatively, such a coupling is fundamentally different
than the gradient coupling mentioned above, since it allows for a
finite coupling between the soft TO mode and the fermions 
even in the $\mathbf{q}\rightarrow 0$ limit. 
Quantitatively,
it is expected that even if the coupling $g$ is small, proximity
to a FE-QCP moves the system towards the strong-coupling regime, since
the pairing interaction becomes singular \citep{Abanov2003}.

\subsection{Superconducting gap equation}

\label{BCS}

Our model consists of Eq. (\ref{eq:H}) supplemented by the bosonic
propagator (\ref{eq:D0}). To obtain the pairing instability, we employ
the standard approach of computing the anomalous fermionic self-energy
via the self-consistent rainbow diagram. To keep the calculation controlled,
we will focus on the BCS (i.e. weak-coupling) regime. While this approximation
does not give access to the behavior at the FE-QCP, it does provide
important insight into the pairing problem as the QCP is approached,
which is our goal in this paper. More specifically, as we show below,
the pairing interaction goes approximately as $g^{2}/\omega_{T}^{2}$
in this approximation. This defines a regime around the QCP where
the soft-mode $\omega_{T}$ remains larger than $g$ such that the
pairing interaction remains in the weak-coupling regime. Thus, the
smaller $g$ is, the closer to the FE-QCP our approach is valid.

\begin{figure}
\centering \includegraphics[width=0.99\linewidth]{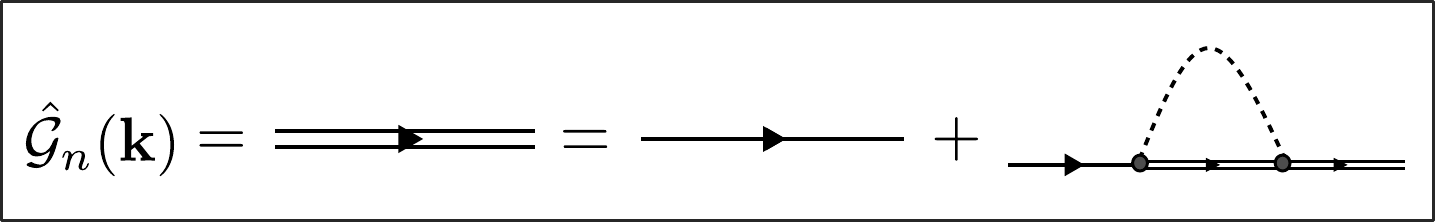}
\caption{Diagrammatic representation of the dressed fermionic Green's function
(double solid lines). The single solid lines correspond to the bare
fermionic Green's function defined in Eq.(\ref{eq:G0}), while the
single dashed line represent the bare bosonic propagator defined in
Eq.(\ref{eq:D0}). The small dark circles denote the interaction vertex
defined in Eq.(\ref{eq:W}).}
\label{fig:diagram} 
\end{figure}

To proceed, we introduce the extended Nambu spinor $\hat{\psi}_{\mathbf{k}}^{\dag}=(c_{\mathbf{k}\uparrow}^{\dag}c_{\mathbf{k}\downarrow}^{\dag}c_{-\mathbf{k}\uparrow}^{\null}c_{-\mathbf{k}\downarrow}^{\null})$
and rewrite the bare electronic Green's function, 
\begin{equation}
\hat{\mathcal{G}}_{0}^{-1}(\mathbf{k},\omega_{n})=\frac{1}{2}\left(i\omega_{n}\sigma_{0}\tau_{0}-\xi_{\mathbf{k}}\sigma_{0}\tau_{3}\right)\text{ ,}\label{eq:G0}
\end{equation}

\noindent where $\omega_{n}=(2n+1)\pi T$ (with $n\in\mathbb{Z}$)
is the fermionic Matsubara frequency and $\sigma_{j}$ ($\tau_{j}$)
denotes the Pauli matrices in spin (particle-hole) space. Moreover,
to shorten the notation we define $\hat{\mathcal{G}}_{0}(\mathbf{k},\omega_{n})\equiv\hat{\mathcal{G}}_{0,n}(\mathbf{k})$,
$\chi_{ij}(\mathbf{q},\Omega_{n})\equiv\chi_{ij,n}(\mathbf{q})$,
and similarly for all the other Green's functions and self-energies
introduced in this section. The coupling between the electrons and
the FE fluctuations dresses the bare electronic propagator according
to Dyson's equation 
\begin{equation}
\hat{\mathcal{G}}_{n}^{-1}(\mathbf{k})=\hat{\mathcal{G}}_{0,n}^{-1}(\mathbf{k})-\hat{\Sigma}_{n}(\mathbf{k})\text{ ,}\label{eq:Dyson}
\end{equation}

\noindent where the fermionic self-energy due to the summation of
rainbow-type diagrams, illustrated in Fig. \ref{fig:diagram}, takes
the form 
\begin{equation}
\hat{\Sigma}_{n}(\mathbf{k})\!=2T\!\!\!\!\sum\limits _{\mathbf{k'},n',ij}\!\!\!\!\chi_{ij,n}\left(\mathbf{k}-\mathbf{k'}\right)\hat{F}_{j}(\mathbf{k},\mathbf{k'}-\mathbf{k})\hat{\mathcal{G}}_{n'}(\mathbf{k'})\hat{F}_{i}(\mathbf{k},\mathbf{k'}-\mathbf{k})\text{ .}\label{eq:SelfE}
\end{equation}

Here, we defined $\hat{F}_{i}(\mathbf{k},\mathbf{k'}-\mathbf{k})\equiv g\left(\hat{\Gamma}(\mathbf{k'})+\hat{\Gamma}(\mathbf{k})\right)/4$, where 

\noindent 
\begin{equation}
\hat{\Gamma}\left(\mathbf{k}\right)=\left(\begin{array}{cccc}
\boldsymbol{\Gamma}_{\uparrow\uparrow}\left(\mathbf{k}\right) & \boldsymbol{\Gamma}_{\uparrow\downarrow}\left(\mathbf{k}\right) & 0 & 0\\
\boldsymbol{\Gamma}_{\downarrow\uparrow}\left(\mathbf{k}\right) & \boldsymbol{\Gamma}_{\downarrow\downarrow}\left(\mathbf{k}\right) & 0 & 0\\
0 & 0 & \boldsymbol{\Gamma}_{\uparrow\uparrow}\left(\mathbf{k}\right) & \boldsymbol{\Gamma}_{\downarrow\uparrow}\left(\mathbf{k}\right)\\
0 & 0 & \boldsymbol{\Gamma}_{\uparrow\downarrow}\left(\mathbf{k}\right) & \boldsymbol{\Gamma}_{\downarrow\downarrow}\left(\mathbf{k}\right)
\end{array}\right)\label{eq:W}
\end{equation}

\noindent denotes the matrix representation of Eq. (\ref{eq:gamma})
in Nambu space.

Note that, in the weak-coupling regime that we study in this paper,
we can neglect the renormalization of the bosonic propagator Eq. (\ref{eq:D0})
caused by the coupling to the electrons. Because of the structure
of the form factor in Eq. (\ref{eq:form_factor}), such a coupling
is expected to give rise to a momentum-dependent Landau damping, which
should affect the pairing problem at the FE-QCP, similarly to the
case of a ferromagnetic QCP \citep{Roussev01}.

We solve the self-consistent equations~\eqref{eq:Dyson} and \eqref{eq:SelfE}
within the weak-coupling approximation, in which the dynamics of the
bosonic propagator is neglected (i.e. $\chi_{ij}\left(\mathbf{k}-\mathbf{k'}\right)\equiv\chi_{ij,0}\left(\mathbf{k}-\mathbf{k'}\right)$
) in lieu of a cut-off $\omega_{c}$, and the states are assumed to
be at the Fermi surface, i.e. $\mathbf{k}=k_{F}\hat{k}$. In addition,
as relevant for STO, we focus on the even-parity pseudospin-singlet channel,
in which case the anomalous part of the self-energy is given by $\hat{\Sigma}_{\mathrm{an}}=i\sigma_{2}\tau_{1}\Delta$.
The resulting linearized gap-equation obtained from the self-consistent
solution of Eq. (\ref{eq:SelfE}) is given by 
\begin{equation}
\Delta(\hat{k})=\text{log}\left(\frac{\kappa\,\omega_{c}}{T_{c}}\right)\int\frac{d\hat{k}'}{4\pi}\Delta(\hat{k}\,')\lambda(\hat{k},\hat{k}\,')\text{ .}\label{eq:BCSgap}
\end{equation}

Here $\kappa=2e^{\gamma}/\pi\approx1.13$, and we defined the dimensionless
pairing interaction function:

\noindent 
\begin{equation}
\lambda(\hat{k},\hat{k}\,')=\frac{\lambda_{0}}{2}\left[L(\hat{k},\hat{k}\,')+L(\hat{k},-\hat{k}\,')\right]\label{eq:lambdak}
\end{equation}

Note that $\lambda(\hat{k},\hat{k}\,')$ has been expressed explicitly
as an even function of $\hat{k}'$, which is accomplished by using
the fact that $\Delta(-\hat{k}\,')=\Delta(\hat{k}\,')$. In this expression,
we defined the dimensionless electron-FE coupling constant $\lambda_{0}=N_{F}g^{2}E_{T}^{-1}$,
where $N_{F}$ is the density of states at the Fermi level. The solid
angle dependence in the coupling function $L(\hat{k},\hat{k}\,')$
causes the gap function $\Delta(\hat{k})$ to be generically anisotropic
around the Fermi surface. This angular dependence is a result of both
the form factor in Eq.~\eqref{eq:form_factor} and the bosonic propagator
Eq.~\eqref{eq:D0}:

\begin{equation}
L(\hat{k},\hat{k}\,')=E_{T}\sum_{ij}f_{ij}(\hat{k},\hat{k}')\chi_{ij}(\hat{k}-\hat{k}\,')\label{eq:Lk}
\end{equation}
with:

\begin{align}
f_{ij}(\hat{k},\hat{k}') & =\left[\Gamma_{i,\downarrow\uparrow}(\mathbf{k})+\Gamma_{i,\downarrow\uparrow}(\mathbf{k}') \right ]\left[\Gamma_{j,\uparrow\downarrow}(\mathbf{k})+\Gamma_{j,\uparrow\downarrow}(\mathbf{k}') \right ]\nonumber \\
 & -\left[ \Gamma_{i,\uparrow\uparrow}(\mathbf{k})+\Gamma_{i,\uparrow\uparrow}(\mathbf{k}')\right ]\left[\Gamma_{j,\downarrow\downarrow}(\mathbf{k})+\Gamma_{j,\downarrow\downarrow}(\mathbf{k}')\right] \text{ .}
\end{align}

\noindent Note that $f_{ij}$ does not depend on the amplitude of $\mathbf{k}$ and $\mathbf{k}'$, since $\Gamma_{i,\alpha\beta}(\mathbf{k})+\Gamma_{i,\alpha\beta}(\mathbf{k}')=[(\hat{k}+\hat{k}')\times\mathbf{\sigma}_{\alpha\beta}]_{i}$.

Since $\chi_{ij}=\chi_{ji}$ and $f_{ij}=f_{ji}^{*}$, it follows
that only the real part of $f_{ij}$ contributes to the sum. Using
the fact that $\left|\hat{k}+\hat{k}'\right|^{2}=2(1+\hat{k}\cdot\hat{k}')$,
the form factor can be re-expressed in the more convenient form:

\begin{equation}
f_{ij}(\hat{k},\hat{k}')\equiv(1+\hat{k}\cdot\hat{k}')\delta_{ij}-\frac{1}{2}\left(\hat{k}_{i}+\hat{k}'_{i}\right)\left(\hat{k}_{j}+\hat{k}'_{j}\right)\label{eq:fk}
\end{equation}

Finally, it is convenient to rewrite the bosonic propagator $\chi_{ij}$
in terms of its diagonalized form:

\begin{equation}
\chi_{ij}^{-1}=\sum_{a=1,2,3}U_{ia}\tilde{\chi}_{aa}^{-1}U_{aj}^{-1}
\end{equation}

Here, $\tilde{\chi}_{aa}^{-1}\equiv E_{T}^{-1}\varpi_{a}^{2}$ give
the three eigenvalues that correspond to the three phonon dispersions.
Furthermore, $U_{ia}\equiv e_{a}^{i}$ and $U^{-1}=U^{T}$, where
$\hat{e}_{a}(\hat{q})$ are the phonon polarizations. Substituting
in Eq. (\ref{eq:Lk}) then yields:

\begin{equation}
L(\hat{k},\hat{k}')=E_{T}^{2}\sum\limits _{a=1,2,3}\frac{\Upsilon_{a}(\hat{k},\hat{k}')}{\varpi_{a}^{2}(\hat{k}-\hat{k}\,')}\label{eq:L_final}
\end{equation}
with the modified form factors:

\begin{equation}
\Upsilon_{a}(\hat{k},\hat{k}')\equiv\sum_{i,j}U_{ia}(\hat{k}-\hat{k}')f_{ij}(\hat{k},\hat{k}')U_{aj}^{-1}(\hat{k}-\hat{k}')\label{eq:Upsilon}
\end{equation}

Related gap equations were derived previously to study the general
problem of superconductivity induced by fluctuations of an odd-parity
bosonic field \citep{Sau14,Fu15,KoziiFu15,Martin17}. While those
works focused on the close competition between even-parity and odd-parity
superconducting instabilities, our emphasis here is on the possible
application of this formalism to STO, which is believed to be a singlet
superconducting state. For this reason, and because the triplet instability
was shown to be subleading in the case of the coupling vertex of Eq.
(\ref{eq:gamma}) \citep{KoziiFu15}, in this paper we restrict our
analysis to the singlet pairing state only.

\section{The isotropic system}

\label{sec:isotropic}

As discussed in the introduction, when the cubic anisotropy term $\varepsilon_{c}$
vanishes in Eq.~\eqref{eq:D0}, diagonalization of $\chi_{ij}^{-1}(\mathbf{q},0)$
leads to a doubly-degenerate transverse mode, $\varpi_{T}^{2}(\mathbf{q})=\omega_{T}^{2}+E_{T}^{2}q^{2}$,
and a purely longitudinal mode, $\varpi_{L}^{2}(\mathbf{q})=\omega_{L}^{2}+E_{L}^{2}q^{2}$.
To derive the effective pairing interaction $\lambda(\hat{k},\hat{k}\,')$
in Eq. (\ref{eq:lambdak}), we must compute the form factors $\Upsilon_{a}(\hat{k},\hat{k}')$
that appear in Eq. (\ref{eq:L_final}). In the case of $\varepsilon_{c}=0$,
it is more convenient to directly invert $\chi_{ij}^{-1}(\mathbf{q},0)$,
which gives:

\begin{equation}
\frac{\chi_{ij}\left(\mathbf{q},0\right)}{E_{T}}=\frac{1}{\varpi_{T}^{2}(\mathbf{q})}\left[\delta_{ij}-\left(\frac{\varpi_{L}^{2}(\mathbf{q})-\varpi_{T}^{2}(\mathbf{q})}{\varpi_{L}^{2}(\mathbf{q})}\right)\hat{q}_{i}\hat{q}_{j}\right]
\end{equation}

We can then directly compute Eq. \eqref{eq:Lk}; using the fact that
$q^{2}=2k_{F}^{2}\left(1-\hat{k}\cdot\hat{k}'\right)$ and $\hat{q}_{i}=\frac{\hat{k}_{i}-\hat{k}'_{i}}{\sqrt{2\left(1-\hat{k}\cdot\hat{k}'\right)}}$,
we find for the effective pairing interaction:

\begin{align}
\lambda(x) & =\lambda_{T}(x)+\lambda_{L}(x)\label{eq:lambdax}\\
\lambda_{a}(x) & =\frac{\lambda_{0}}{2}\left(\frac{E_{T}}{\omega_{a}}\right)^{2}\sum_{\pm}\frac{1\pm x}{1+2\eta_{a}^{2}(1\mp x)}\label{eq:lambdaa}
\end{align}
where $\eta_{a}\equiv\frac{E_{a}k_{F}}{\omega_{a}}$ is a dimensionless
parameter, $a=T$ or $a=L$, $\omega_{a}$ are the optical gaps of
each mode at the zone center, and $x=\hat{k}\cdot\hat{k}'=\cos\theta_{k,k'}$
is the relative angle between scattering momenta. Recall that throughout
this work, momentum (and thus $k_{F}$) is dimensionless as it is
expressed in units of $\pi/a$, whereas energies ($E_{a}$, $\varepsilon_{c}$)
and frequencies ($\omega_{a}$) have units of energy.

\begin{figure}
\centering \includegraphics[width=1\linewidth]{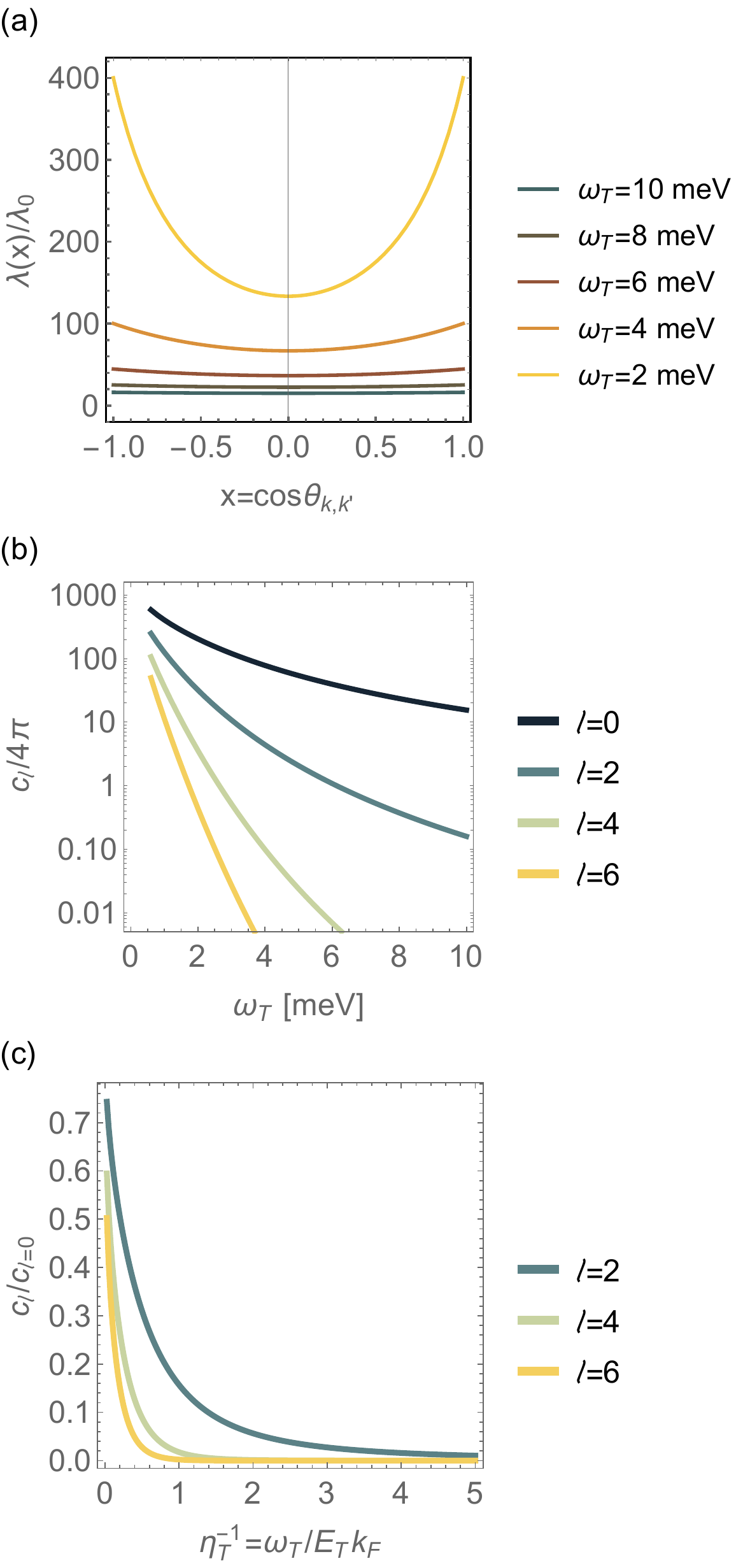}
\caption{(a) Effective pairing interaction $\lambda(x)$, in units of the dimensionless
coupling constant $\lambda_{0}$, as function of the relative angle
between the scattering vectors $x=\cos\theta_{k,k'}$, for different
values of the TO mode frequency $\omega_{T}$. Here, we set $k_{F}=0.05$
and $E_{T}=40$ meV as relevant for STO, yielding $E_{T}k_{F}=2$
meV. (b) Coefficients $c_{l}$ of the $l=0,2,4$ and $6$ pairing
channels as function of $\omega_{T}$. (c) Ratio between the subdominant
coefficients $c_{l>0}$ specified in the legend and the dominant $c_{0}$
coefficient as function of the dimensionless parameter $\frac{1}{\eta_{T}}=\frac{\omega_{T}}{E_{T}k_{F}}$.}
\label{fig:isotropic} 
\end{figure}

The pairing interaction $\lambda(x)$ is shown in Fig.~\ref{fig:isotropic}(a)
for several values of the transverse optical gap $\omega_{T}$. It
is positive for all $x$ and thus, in agreement with Ref. \citep{KoziiFu15},
it provides an attractive interaction in the even-parity pseudospin-singlet
channel. Moreover, because close to the FE instability $\omega_{L}\gg\omega_{T}$,
the contribution of the longitudinal sector to the pairing interaction
$\lambda(x)$ is negligible, $\frac{\lambda_{L}(x)}{\lambda_{T}(x)}\sim\left(\frac{\omega_{T}}{\omega_{L}}\right)^{2}\ll1$.
Consequently, the pairing interaction is mediated primarily through
the coupling to the soft transverse mode. This is an important result
of our work: in contrast to the dipolar-mediated gradient coupling,
the spin-orbit-mediated coupling depends only on the TO mode.

As shown in Fig.~\ref{fig:isotropic}(a), as the system approaches
the FE transition and $\omega_{T}$ decreases, the pairing interaction
$\lambda(x)$ increases and becomes more anisotropic, as it depends
more strongly on $\theta_{k,k'}$. The maximum of the pairing interaction
happens for nearly parallel momenta, $\theta_{k,k'}\approx0$. This
is a direct consequence of the fact that the FE fluctuations are peaked
at zero momentum transfer. For $\theta_{k,k'}\approx0$, $\lambda(x=\pm1)\approx\lambda_{T}(x=\pm1)=\lambda_{0}\left(\frac{E_{T}}{\omega_{T}}\right)^{2}$
grows with the softening of the TO mode, whereas for all other angles,
the pairing interaction is suppressed by the factor $\eta_{T}^{2}=\left(\frac{E_{T}k_{F}}{\omega_{T}}\right)^{2}$
in the denominator of Eq.~\eqref{eq:lambdaa}. As the TO gap $\omega_{T}$
is reduced and becomes comparable to the characteristic energy $E_{T}k_{F}$,
the contribution from $\eta_{T}$ in the denominator becomes important,
and the $\theta_{k,k'}$ anisotropy of the pairing interaction increases,
acquiring a pronounced minimum for perpendicular scattering $x=0$.
As will be shown below, this has important consequences for the competition
of the various even-parity superconducting channels.

In order to solve the gap equation~\eqref{eq:BCSgap}, we exploit
the rotational invariance of the system and expand the effective interaction
and the gap function into spherical harmonics, 
\begin{align}
\lambda(\hat{k},\hat{k}') & =\lambda_{0}\sum_{l,m}c_{l}Y_{l}^{m}(\hat{k})\left[Y_{l}^{m}(\hat{k}')\right]^{*}\\
\Delta(\hat{k}) & =\sum_{l,m}d_{l}Y_{l}^{m}(\hat{k})%
\end{align}

As a result, the linearized gap equation decouples into orthogonal
even-parity channels characterized by the Cooper-pair angular momentum
$l=2n$ ($n\in\mathbb{N}$),
\begin{equation}
1=\frac{c_{l}}{4\pi}\lambda_{0}\log\left(\frac{1.13\omega_{c}}{k_{B}T_{c}^{(l)}}\right).\label{eq:eigval_l}
\end{equation}

The largest coefficient $c_{l}$ gives the largest superconducting
transition temperature $T_{c}^{(l)}$, thus defining the leading superconducting
instability channel. The coefficients $c_{l}$ corresponding to the
four largest even angular momenta $l$ are shown in Fig.~\ref{fig:isotropic}(b)
as function of the TO mode frequency $\omega_{T}$. As expected, all
coefficients grow for decreasing $\omega_{T}$, in agreement with
the increase of the pairing interaction {[}Fig.~\ref{fig:isotropic}(a){]}.
The $l=0$ channel has the largest coefficient, signaling an isotropic
$s$-wave gap function at $T_{c}$, as expected from an overall attractive
pairing interaction $\lambda(x)$. Moreover, as shown in Fig.~\ref{fig:isotropic}(b),
the coefficients of the subleading even-parity channels $l=2$, $l=4$
and $l=6$ grow faster than the isotropic $l=0$ solution as the system
approaches the FE instability, although $c_{0}$ always remains larger
than $c_{l>0}$ in our calcultions.

To understand this behavior, let us focus again on the effective pairing
interaction $\lambda(x)$ in Eq.~\eqref{eq:lambdaa}. When the frequency
of the TO mode is sufficiently large such that $\eta_{T}=\frac{E_{T}k_{F}}{\omega_{T}}\ll1$,
the interaction is essentially constant, $\lambda(x)\simeq\lambda_{0}\frac{E_{T}^{2}}{\omega_{T}^{2}}$.
As a result, $c_{0}\gg c_{l>0}$, in analogy to the standard case
of a phonon-mediated pairing interaction, which is momentum-independent.
Note that this form of the interaction was proposed on phenomenological
grounds in Ref. \citep{Edge15}. As the TO mode becomes softer, however,
$\eta_{T}$ grows and the interaction acquires a strong $\theta_{k,k'}$
anisotropy {[}Fig.~\ref{fig:isotropic}(a){]}, becoming strongly
peaked near zero momentum transfer. As a result, the coefficients
of the higher-order harmonics increase and approach the isotropic
coefficient, $c_{l>0}\rightarrow c_{0}$, in the regime $\omega_{T}\ll E_{T}k_{F}$,
as illustrated in Fig.~\ref{fig:isotropic}(c). This behavior is
reminiscent of the case of pairing mediated by nematic fluctuations,
which favor all pairing channels due to the fact that they are also
strongly peaked at zero-momentum \citep{Lederer15,Kang16,Klein18}.

\section{The cubic system}

\label{sec:cubic}

In an actual cubic material, the symmetry is lowered from the continuous
rotation group to the discrete point group $O_{h}$. In our model,
the cubic symmetry of the lattice is manifested by the cubic anisotropy
term $\varepsilon_{c}$ in the bosonic propagator (\ref{eq:D0}).
In this section, we investigate the impact of this term on the pairing
state promoted by the FE fluctuations. Clearly, because $\varepsilon_{c}$
lowers the symmetry of the effective interaction Eq.~\eqref{eq:Lk},
the superconducting gap solution $\Delta(\hat{k})$ will no longer be isotropic.
As we will show, the anisotropy of the gap generated by this term
becomes significant near the FE transition. Note that we neglect,
for simplicity, the effect of the cubic crystal field on the electronic
band dispersion Eq.~\eqref{eq:G0}.

As explained in the introduction and also discussed in Ref. \citep{Woelfle18},
the main effect of the cubic anisotropy on the bosonic degrees of
freedom is to couple the polarization of the transverse and longitudinal
phonon modes~\citep{Ruhman19}. While the anisotropic term is not
small, $\varepsilon_{c} \sim E_T$ according to fits to neutron data~\citep{Yamada69}, treating $\varepsilon_{c}$
as a perturbation can give powerful insight into the problem. 
Our
strategy in this section is thus to get insight of the gap solution from the perturbative expansion
of the pairing interaction in $\varepsilon_{c}$, and then compare
the results with the numerical solution of the gap equation \eqref{eq:BCSgap}.

Treating $\varepsilon_{c}$ as a perturbation, we find that the leading-order
correction arising from the polarization of the modes of $\mathcal{\chi}_{ij}(\hat{k}-\hat{k}')$
together with the form factor $f_{ij}(\hat{k},\hat{k}')$ in Eq.~\eqref{eq:Lk}
vanishes (see Appendix~\ref{App:perturbation} for details). Thus,
the effective pairing interaction is altered solely by the first order
correction of the dispersion of the modes. The corresponding expression
for the modified interaction Eq.~\eqref{eq:Lk} reads, 
\begin{align}
L(\hat{k}-\hat{k}\,') & \simeq\left(\frac{E_{T}}{\omega_{a}}\right)^{2}\sum_{a=T,L}\frac{1+x}{1+2\eta_{a}^{2}(1-x)}\times\label{eq:Lkpert}\\
 & \left[1-\frac{\varepsilon_{c}^{2}k_{F}^{2}}{2(1-x)\omega_{a}^{2}\left[1+2\eta_{a}^{2}(1-x)\right]}\zeta_{a}(\hat{k}-\hat{k}')\right]\nonumber 
\end{align}
which is no longer rotational invariant, i.e. it no longer depends
only on the relative angle between $\hat{k}$ and $\hat{k}'$ ($x=\cos\hat{k}\cdot\hat{k}'$),
but acquires a cubic angular dependence through the form factors $\zeta_{T}$
and $\zeta_{L}$ defined as:
\begin{align}
\zeta_{T}(\hat{q}) & =2\left(\hat{q}_{x}^{2}\hat{q}_{y}^{2}+\hat{q}_{x}^{2}\hat{q}_{z}^{2}+\hat{q}_{y}^{2}\hat{q}_{z}^{2}\right)\label{eq:zetaT}\\
\zeta_{L}(\hat{q}) & =\hat{q}_{x}^{4}+\hat{q}_{y}^{4}+\hat{q}_{z}^{4}.\label{eq:zetaL}
\end{align}

\begin{figure}
\centering \includegraphics[width=1\linewidth]{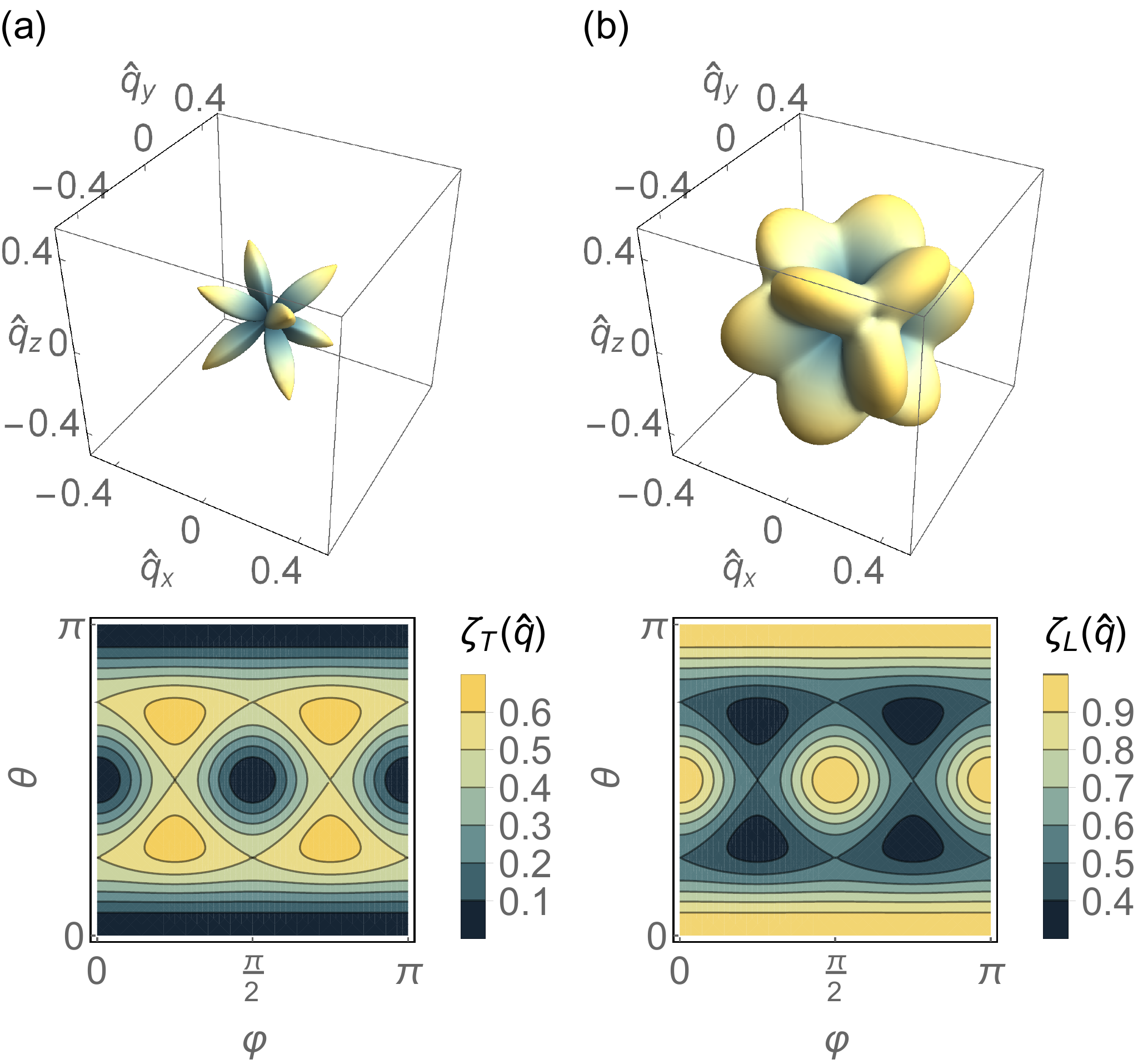}
\caption{Form factors $\zeta_{a}(\hat{q})$ obtained in the perturbative treatment
of the pairing interaction to leading order in the cubic anisotropy
term $\varepsilon_{c}$. (a) $\zeta_{T}(\hat{q})$ {[}Eq.~\eqref{eq:zetaT}{]}
and (b) $\zeta_{L}(\hat{q})$ {[}Eq.~\eqref{eq:zetaL}{]}. These
form factors enter the effective pairing interaction $L(\hat{q})$
given by Eq.~\eqref{eq:Lkpert}.}
\label{fig:zeta}
\end{figure}

These form factors are plotted in Fig.~\ref{fig:zeta} to highlight
their cubic symmetry. The final perturbative expression for the pairing
interaction is then $\lambda(\hat{k},\hat{k}')\simeq\lambda(x)+\lambda_{c}(\hat{k},\hat{k}')$,
with contributions from the transverse and longitudinal sectors $\lambda_{c}(\hat{k},\hat{k}')=\lambda_{c,T}(\hat{k},\hat{k}')+\lambda_{c,L}(\hat{k},\hat{k}')$:
\begin{widetext}
\begin{align}
\lambda_{c,T}(\hat{k},\hat{k}') & =-\frac{\lambda_{0}}{2}\left(\frac{E_{T}}{\omega_{T}}\right)^{2}\left(\frac{\varepsilon_{c}k_{F}}{\omega_{T}}\right)^{2}\left[\frac{1+x}{1-x}\frac{\sum_{i>j}(\hat{k}_{i}-\hat{k}'_{i})^{2}(\hat{k}_{j}-\hat{k}'_{j})^{2}}{\left[1+2\eta_{T}^{2}(1-x)\right]^{2}}+\frac{1-x}{1+x}\frac{\sum_{i>j}(\hat{k}_{i}+\hat{k}'_{i})^{2}(\hat{k}_{j}+\hat{k}'_{j})^{2}}{\left[1+2\eta_{T}^{2}(1+x)\right]^{2}}\right]\label{eq:lambdacT}\\
\lambda_{c,L}(\hat{k},\hat{k}') & =-\frac{\lambda_{0}}{4}\left(\frac{E_{T}}{\omega_{L}}\right)^{2}\left(\frac{\varepsilon_{c}k_{F}}{\omega_{L}}\right)^{2}\left[\frac{1+x}{1-x}\frac{\sum_{i}(\hat{k}_{i}-\hat{k}_{i}')^{4}}{\left[1+2\eta_{L}^{2}(1-x)\right]^{2}}+\frac{1-x}{1+x}\frac{\sum_{i}(\hat{k}_{i}+\hat{k}_{i}')^{4}}{\left[1+2\eta_{L}^{2}(1-x)\right]^{2}}\right].\label{eq:lambdacL}
\end{align}
\end{widetext}

The main point of Eqs.~\eqref{eq:zetaT}-\eqref{eq:lambdacL} is to illustrate that the gap function is no longer isotropic, but carries on the information encoded in $\varepsilon_c$ about the cubic anisotropy of the system. These expression will be used later on to understand qualitatively the results of this section.

In analogy to our solution for the isotropic case,
we now have to expand the full pairing 
interaction~\eqref{eq:lambdak} (i.e. without assuming small $\varepsilon_c$) into an appropriate set of functions. Instead of spherical
harmonics, we use the cubic harmonics $K_{a}^{\Gamma}(\hat{k})$ for
each irreducible representation (irrep) $\Gamma$ of the $O_{h}$
cubic group~\citep{Altmann65}. For a given irrep, the
basis functions $K_{a}^{\Gamma}(\hat{k})$ can be expressed in terms
of spherical harmonics $Y_{a}^{m}$ corresponding to different values
of the angular momentum $a$. The set of basis functions with smallest
angular momentum projection $a$ of the five even-parity irreps ($A_{1g}$,
$E_{g}$, $T_{2g}$, $T_{1g}$ and $A_{2g}$) are shown in Table~\ref{tab:Ki}.
Note that the $A_{1g}$ irrep basis contains functions associated
not only with zero angular momentum (``$s$-wave''), $K_{0}^{A_{1g}}$,
but also with $a=4$ angular momentum (``$g$-wave''), $K_{4}^{A_{1g}}$.
This is not surprising, since angular momentum is not a good quantum
number for the cubic system.

The full pairing interaction $\lambda(\hat{k},\hat{k}')$
(and of course the perturbative expressions Eqs.~\eqref{eq:lambdacT}-\eqref{eq:lambdacL}) mixes, within
each irrep $\Gamma$, the set of functions $K_{a}^{\Gamma}(\hat{k})$.
It can therefore be expanded in the following form 
\begin{equation}
\lambda(\hat{k},\hat{k}')=\lambda_{0}\sum_{\Gamma}\sum_{ab}c_{ab}^{\Gamma}K_{a}^{\Gamma}(\hat{k})K_{b}^{\Gamma}(\hat{k}')\label{eq:lambda_exp}
\end{equation}
where the coefficients $c_{ab}$ are real numbers. The presence of
non-zero $c_{a\neq b}^{\Gamma}\neq0$ implies mixing of angular momenta,
and therefore the solution of the gap function $\Delta(\hat{k})$
will be in general a combination of cubic harmonics with different
angular momenta. We show explicitly in Appendix~\ref{App:c40} that
$c_{04}^{A_{1g}}\neq0$ follows from the perturbative expressions
Eqs.~\eqref{eq:lambdacT}-\eqref{eq:lambdacL}.

To proceed, we expand the gap function in cubic harmonics $\Delta(\hat{k})=\sum_{\Gamma}\sum_{a}b_{a}^{\Gamma}K_{a}^{\Gamma}(\hat{k})$.
As a result, the superconducting gap equation \eqref{eq:BCSgap} is
decoupled into different irreducible representation channels $\Gamma$:
\begin{equation}
\sum_{a}d_{a}^{\Gamma}K_{a}^{\Gamma}(\hat{k})=\lambda_{0}\log\left(\frac{1.13\omega_{c}}{k_{B}T_{c}}\right)\frac{1}{4\pi}\sum_{a,b}d_{a}^{\Gamma}c_{ab}^{\Gamma}K_{b}^{\Gamma}(\hat{k}).\label{eq:gap_eq_cubic}
\end{equation}

Our task is then reduced to compute the pairing interaction matrix
elements $c_{ab}^{\Gamma}$ in Eq.~\eqref{eq:lambda_exp}, which
is numerically straightforward, and then obtain the corresponding
largest eigenvalue for each channel $\Gamma$. To make the calculations
analytically tractable, we truncate the expansion in Eq.~\eqref{eq:lambda_exp}
into a finite-dimensional matrix in the $a,b$ angular momentum subspace.
This is justified as long as the coefficients $c_{ab}^{\Gamma}$ decrease
with increasing $a,b$, which we show to be the case below. Note that
such a truncation is analogous to the leading angular harmonics approximation
(LAHA) method employed to study the gap functions of iron-based superconductors
\citep{Maiti11}.

\begin{table}
\centering{}\caption{Even-parity basis gap-functions of the $O_{h}$ cubic group for each
irreducible representation $\Gamma$. Each function is given in terms
of real spherical harmonics $Y_{l}^{m,c}(\hat{k})\equiv\frac{1}{\sqrt{2}}\left(Y_{l}^{-m}(\hat{k})+Y_{l}^{m}(\hat{k})\right)$
and $Y_{l}^{m,s}(\hat{k})\equiv\frac{i}{\sqrt{2}}\left(Y_{l}^{-m}(\hat{k})-Y_{l}^{m}(\hat{k})\right)$.
For multi-dimensional irreps, only one of the orthogonal basis is
given. From Ref.~\citep{Altmann65}.}
\label{tab:Ki} %
\begin{tabular}{|l|l|}
\hline 
\textbf{Irrep $\Gamma$}  & \textit{Representative basis functions}\tabularnewline
\hline 
\multirow{2}{*}{$A_{1g}$} & $K_{0}^{A_{1g}}=Y_{0}^{0}$ \tabularnewline
 & $K_{4}^{A_{1g}}(\hat{k})=\frac{1}{2}\sqrt{\frac{7}{3}}Y_{4}^{0}(\hat{k})+\frac{1}{2}\sqrt{\frac{5}{3}}Y_{4}^{4,c}(\hat{k})$ \tabularnewline
\hline 
\multirow{2}{*}{$E_{g}$} & $K_{2}^{E_{g}}(\hat{k})=Y_{2}^{2,c}(\hat{k})$\tabularnewline
 & $K_{4}^{E_{g}}(\hat{k})=Y_{4}^{2,c}(\hat{k})$\tabularnewline
\hline 
\multirow{2}{*}{$T_{2g}$} & $K_{2}^{T_{2g}}(\hat{k})=Y_{2}^{2,s}(\hat{k})$ \tabularnewline
 & $K_{4}^{T_{2g}}(\hat{k})=-Y_{4}^{2,s}(\hat{k})$ \tabularnewline
\hline 
\multirow{2}{*}{$T_{1g}$} & $K_{4}^{T_{1g}}(\hat{k})=Y_{4}^{4,s}(\hat{k})$ \tabularnewline
 & $K_{6}^{T_{1g}}(\hat{k})=Y_{6}^{4,s}(\hat{k})$\tabularnewline
\hline 
\multirow{2}{*}{$A_{2g}$} & $K_{6}^{A_{2g}}(\hat{k})=\frac{\sqrt{11}}{4}Y_{6}^{2,c}(\hat{k})-\frac{\sqrt{5}}{4}Y_{6}^{6,c}(\hat{k})$\tabularnewline
 & $K_{10}^{A_{2g}}(\hat{k})=0.8Y_{10}^{2,c}(\hat{k})+0.157Y_{10}^{6,c}(\hat{k})-0.576Y_{10}^{10,c}(\hat{k})$\tabularnewline
\hline 
\end{tabular}
\end{table}

We focus first on the $A_{1g}$ pairing channel of Eq.~\eqref{eq:gap_eq_cubic},
corresponding to the trivial irrep of the $O_{h}$ group. In the previous
section, without the cubic anisotropy ($\varepsilon_{c}=0$), the
leading gap function was found to be the isotropic one, i.e. $Y_{0}^{0}$
($s$-wave). Truncating the pairing interaction expansion \eqref{eq:lambda_exp}
to second-order gives:
\begin{equation}
\lambda(\hat{k},\hat{k}')=\lambda_{0}\begin{pmatrix}K_{0} & K_{4}(\hat{k})\end{pmatrix}\begin{pmatrix}c_{00} & c_{04}\\
c_{04} & c_{44}
\end{pmatrix}\begin{pmatrix}K_{0}\\
K_{4}(\hat{k}')
\end{pmatrix}\label{eq:aux_lambda}
\end{equation}

Here, the $\Gamma=A_{1g}$ superscript has been dropped for clarity.
Solution of the gap equation gives:
\begin{align}
1 & =\lambda_{0}\frac{\alpha}{4\pi}\log\left(\frac{1.13\omega_{c}}{k_{B}T_{c}}\right)\label{eq:eigval_A1g}\\
\Delta(\hat{k}) & =K_{0}+d_{4}K_{4}(\hat{k}),\label{eq:Delta_A1g}
\end{align}
where $\alpha$ is largest eigenvalue of the matrix in Eq. \eqref{eq:aux_lambda}
and $d_{4}$ can be obtained from the corresponding eigenvector: 
\begin{align}
\alpha & =\frac{1}{2}\left(c_{00}+c_{44}+\sqrt{(c_{00}-c_{44})^{2}+4c_{04}^{2}}\right)\label{eq:alpha}\\
d_{4} & =\frac{2c_{04}}{c_{00}-c_{44}+\sqrt{(c_{00}-c_{44})^{2}+4c_{04}^{2}}}.\label{eq:d4}
\end{align}

The $A_{1g}$ gap function $\Delta(\hat{k})$ thus acquires an anisotropic
gap structure due to the cubic harmonic $K_{4}$ (see Table~\ref{tab:Ki}).
To see why the appearance of this harmonic is generally expected,
we rewrite it in Cartesian coordinates:
\begin{equation}
K_{4}(\hat{k})=\sqrt{\frac{21}{16\pi}}\left[\hat{k}_{x}^{4}+\hat{k}_{y}^{4}+\hat{k}_{z}^{4}-3\left(\hat{k}_{x}^{2}\hat{k}_{y}^{2}+\hat{k}_{x}^{2}\hat{k}_{z}^{2}+\hat{k}_{y}^{2}\hat{k}_{z}^{2}\right)\right]\label{eq:K4}
\end{equation}
Comparing to the form factors $\zeta_{a}(\hat{k})$ introduced perturbatively by
the cubic anisotropy term $\varepsilon_{c}$ in Eqs. \eqref{eq:zetaT}
and \eqref{eq:zetaL}, it is clear that $K_{4}(\hat{k})\propto\zeta_{L}(\hat{k})-\frac{3}{2}\,\zeta_{T}(\hat{k})$.
Thus, this form of the gap anisotropy simply reflects the anisotropy
in the bosonic propagator.

\begin{figure}
\centering \includegraphics[width=1\linewidth]{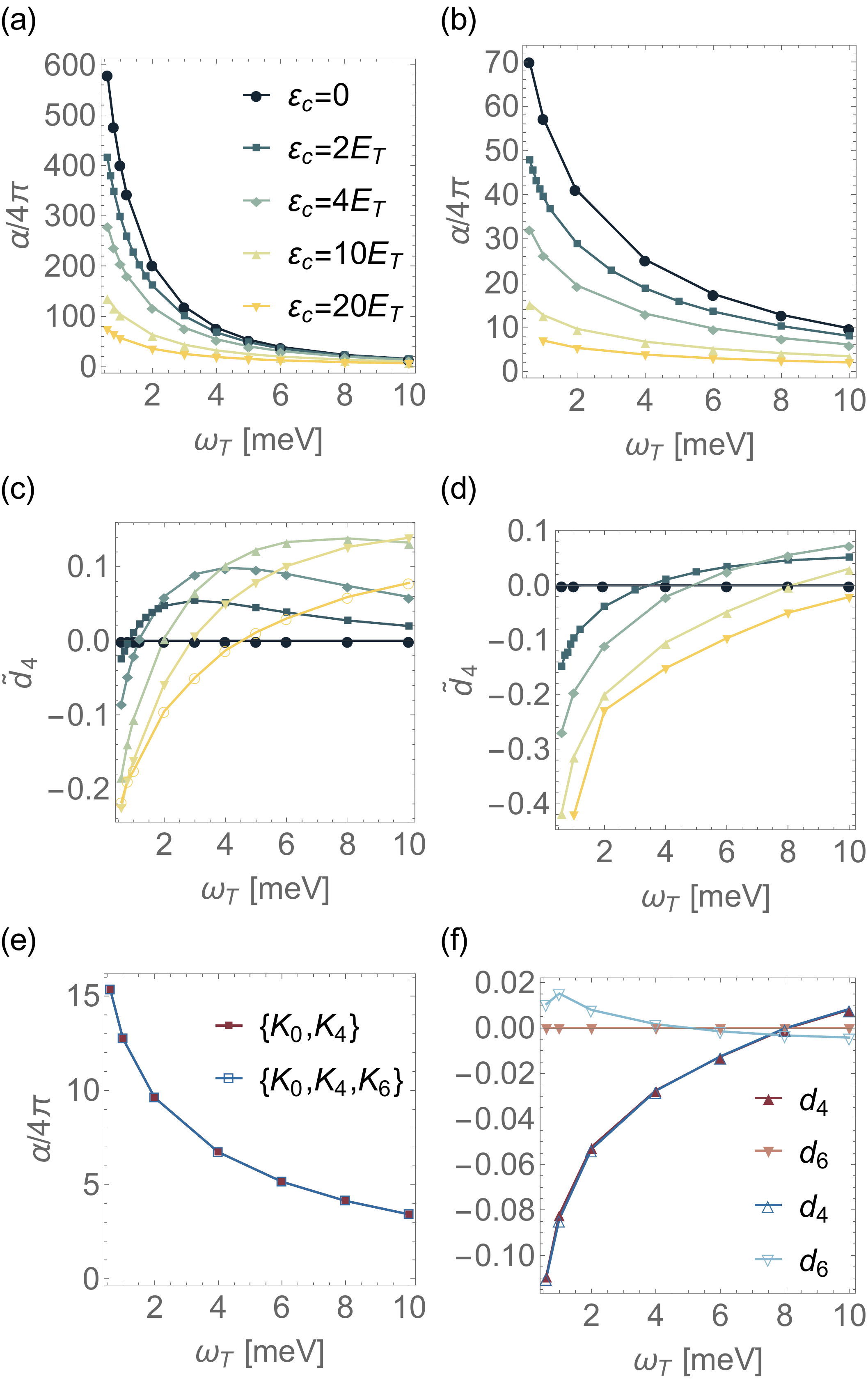}
\caption{Eigenvalue of the $A_{1g}$ pairing channel $\alpha$, Eq.~\eqref{eq:alpha},
for various values of the cubic anisotropy $\varepsilon_{c}$ as specified
in the legend, and for (a) $k_{F}=0.05$ and (b) $k_{F}=0.2$. The
corresponding renormalized coefficient $\tilde{d}_{4}$, which is
proportional to the anisotropy of the gap function (see definition
in the main text), is shown in panels (c) and (d), respectively. (e) Comparison of the computed eigenvalue $\alpha$ for the $\varepsilon_c=10 E_T$ case in panel (b) with $k_F=0.2$ by including also the next cubic harmonic $K_6(\hat{k})$ in the truncation Eq.~\eqref{eq:aux_lambda}. (f) Coefficients of the eigenvector $\Delta(\hat{k})=K_0+d_4 K_4(\hat{k})+d_6 K_6(\hat{k})$ of the eigenvalue cases in panel (e). }
\label{fig:A1g_d4} 
\end{figure}

Fig.~\ref{fig:A1g_d4}(a) shows the behavior of the coefficient $\alpha$
as the system approaches the FE instability for various values of
the cubic crystal field $\varepsilon_{c}$. In agreement with what
we found in the absence of cubic anisotropy ($\varepsilon_{c}=0$,
dark blue curve in Fig.~\ref{fig:A1g_d4}(a)), the eigenvalue $\alpha$
grows as $\omega_{T}\rightarrow0$ for all $\varepsilon_{c}$ values,
implying an enhancement of $T_{c}$ as the FE transition is approached.
Moreover, $\alpha$ decreases non-monotonically with the cubic anisotropy
$\varepsilon_{c}$ in all the cases we studied (not shown). A larger Fermi momentum $k_{F}$
also suppresses the eigenvalue $\alpha$, as seen by comparing Figs.~\ref{fig:A1g_d4}(a)
and (b), which correspond to $k_{F}=0.05$ and $k_{F}=0.2$, respectively. Note that the dimensionless electron-FE
coupling $\lambda_{0}=N_{F}g^{2}/E_{T}$ may also be enhanced by increasing
$k_{F}$, particularly if the system starts in the very dilute regime.
We do not include this effect in our calculation.  Finally, to show that the second-order
truncation in Eq. \eqref{eq:aux_lambda} is enough to correctly capture
the pairing potential, in Fig.~\ref{fig:A1g_d4}(e) we compare the $A_{1g}$ eigenvalue $\alpha$ obtained by truncating at $K_4(\hat{k})$ (filled blue squares) and to next order $K_6(\hat{k})$ (empty yellow squares). Clearly, the correction is very small. Indeed, the coefficient of the next cubic harmonic of the eigenvector $\Delta(\hat{k})=K_0+d_4 K_4(\hat{k})+d_6 K_6(\hat{k})$ is still significantly smaller in this regime of parameter space, i.e. $d_6 \ll d_4$, illustrated in Fig.~\ref{fig:A1g_d4}(f).

Fig.~\ref{fig:A1g_d4}(c) shows the coefficient $d_{4}$ calculated
from Eq. \eqref{eq:d4}, which appears in front of the anisotropic
contribution $K_{4}(\hat{k})$ to the gap function, $\Delta(\hat{k})=K_{0}+d_{4}K_{4}(\hat{k}).$
Since the absolute value of the gap is not fixed by the linearized
gap equations, we plot $\tilde{d}_{4}=cd_{4}$, where $c=3.82$ is the
peak-to-peak amplitude of the anisotropic function $K_{4}(\hat{k})/K_{0}$,
i.e. the difference between the maximum and the minimum of this function. 
As a result, $\tilde{d}_{4}$ gives the relative anisotropy of the
gap function. The curves shown in Figs.~\ref{fig:A1g_d4}(c)-(d)
correspond to the same $\varepsilon_{c}$ values in Figs.~\ref{fig:A1g_d4}(a)-(b).
As expected, the gap anisotropy $\tilde{d}_{4}$ increases with increasing
cubic anisotropy $\varepsilon_{c}$. Interestingly, for a fixed $\varepsilon_{c}$
value, the gap anisotropy shows a non-monotonic behavior for decreasing
$\omega_{T}$, even changing sign below a critical value of the TO
mode frequency $\omega_{T}^{*}$. This critical value depends not
only on the cubic anisotropy $\varepsilon_{c}$, but also on the value
of $k_{F}$, as it can be seen by comparing panels \ref{fig:A1g_d4}(c)
and (d). For the parameters explored here, we find the biggest gap
anisotropy to be around $40\%$, taking place at large cubic anisotropies
and small values of the TO gap $\omega_{T}$.

\begin{figure}
\centering \includegraphics[width=1\linewidth]{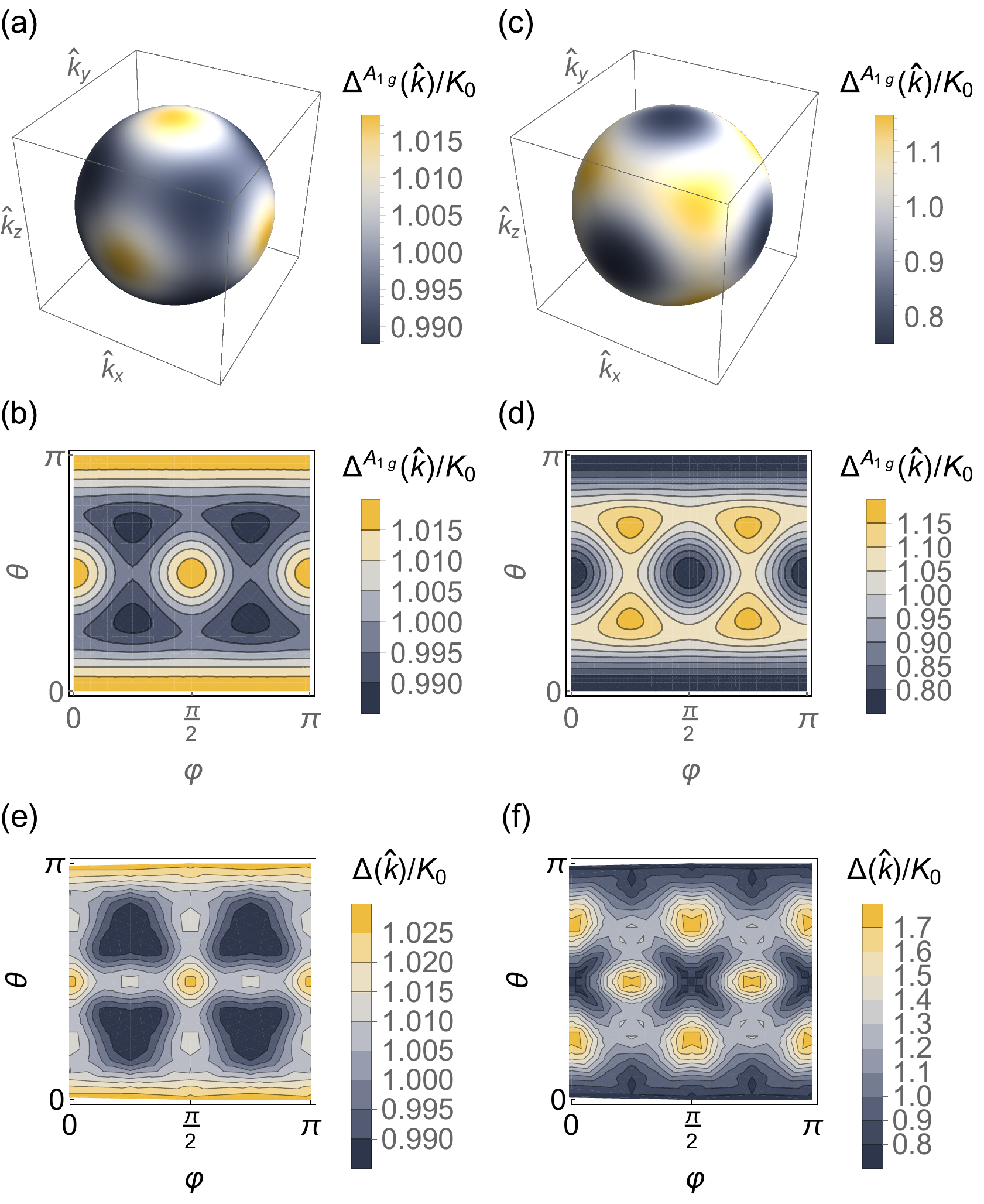}
\caption{Gap function associated with the $A_{1g}$ pairing instability projected
onto the Fermi surface for (a)-(c) $\omega_{T}=10$ meV and (d)-(f)
$\omega_{T}=0.6$ meV. The (a) and (d) ((b) and (e)) panels are three-dimensional
(two-dimensional) representations in a sphere (projected on the $\theta-\varphi$
plane). Panels (c) and (f) show the projected gap function of the highest eigenvalue obtained from the direct
numerical solution of the gap equation \eqref{eq:BCSgap}. In all panels, $E_{T}=40$ meV, $E_{L}=E_{T}/10$, $\omega_{L}=100$ meV, $\varepsilon_{c}=10E_{T}$ and $k_{F}=0.2$.} 
\label{fig:A1g_structure}
\end{figure}

The full angular dependence of the $A_{1g}$ gap function for $\varepsilon_{c}=10E_{T}$
and $k_{F}=0.2$ is shown in Fig.~\ref{fig:A1g_structure} for a
large and a small value of $\omega_{T}$. In Figs.~\ref{fig:A1g_structure}(a)-(b),
because the TO frequency $\omega_{T}=10$ meV is only slightly above
the critical value $\omega_{T}^{*}$ for which $d_{4}$ changes sign,
the anisotropy of $\Delta(\hat{k})$ is very small. Moreover, because
$d_{4}>0$ {[}see Fig.~\ref{fig:A1g_d4}(d){]}, the gap maxima (light
yellow) are located along the $[100]$ and symmetry-related directions,
whereas the gap minima (dark blue) appear along the diagonal $[111]$
and symmetry-related directions. For $\omega_{T}<\omega_{T}^{*}$,
the gap anisotropy is reversed, since $d_{4}<0$. This is illustrated
in Figs.~\ref{fig:A1g_structure}(d)-(e), obtained for $\omega_{T}=0.6$
meV. Besides the switching between the positions of the minima and
maxima, we note that the magnitude of the anisotropy is also enhanced,
as expected from the enhancement of the magnitude of $d_{4}$ upon
decreasing $\omega_{T}$.

In the figures discussed above, we relied on a finite truncation in the expansion of the full interaction Eq.~\eqref{eq:lambda_exp}. To check whether the conclusions obtained from this
method hold, 
we numerically solved the integral equation Eq.~\eqref{eq:BCSgap}
by using a Lebedev quadrature on the sphere~\citep{Lebedev76}, an
optimized method for cubic numerical integration. As shown in Figs.~
\ref{fig:A1g_structure}(e)-(f), the numerical gap agrees relatively well
with the 
gap obtained by the truncation method, shown in Fig.~ \ref{fig:A1g_structure}(b) and Fig.~ \ref{fig:A1g_structure}(d).
This agreement includes the main conclusions that the gap anisotropy
changes sign and enhances in magnitude as $\omega_{T}$ decreases.
Note that the quantitative agreement is better for larger
values of $\omega_{T}$, indicating that higher-order harmonics become more important as the FE-QCP is approached. This is consistent with what we found in Fig. \ref{fig:A1g_d4}(f).

\begin{figure}
\centering \includegraphics[width=1\linewidth]{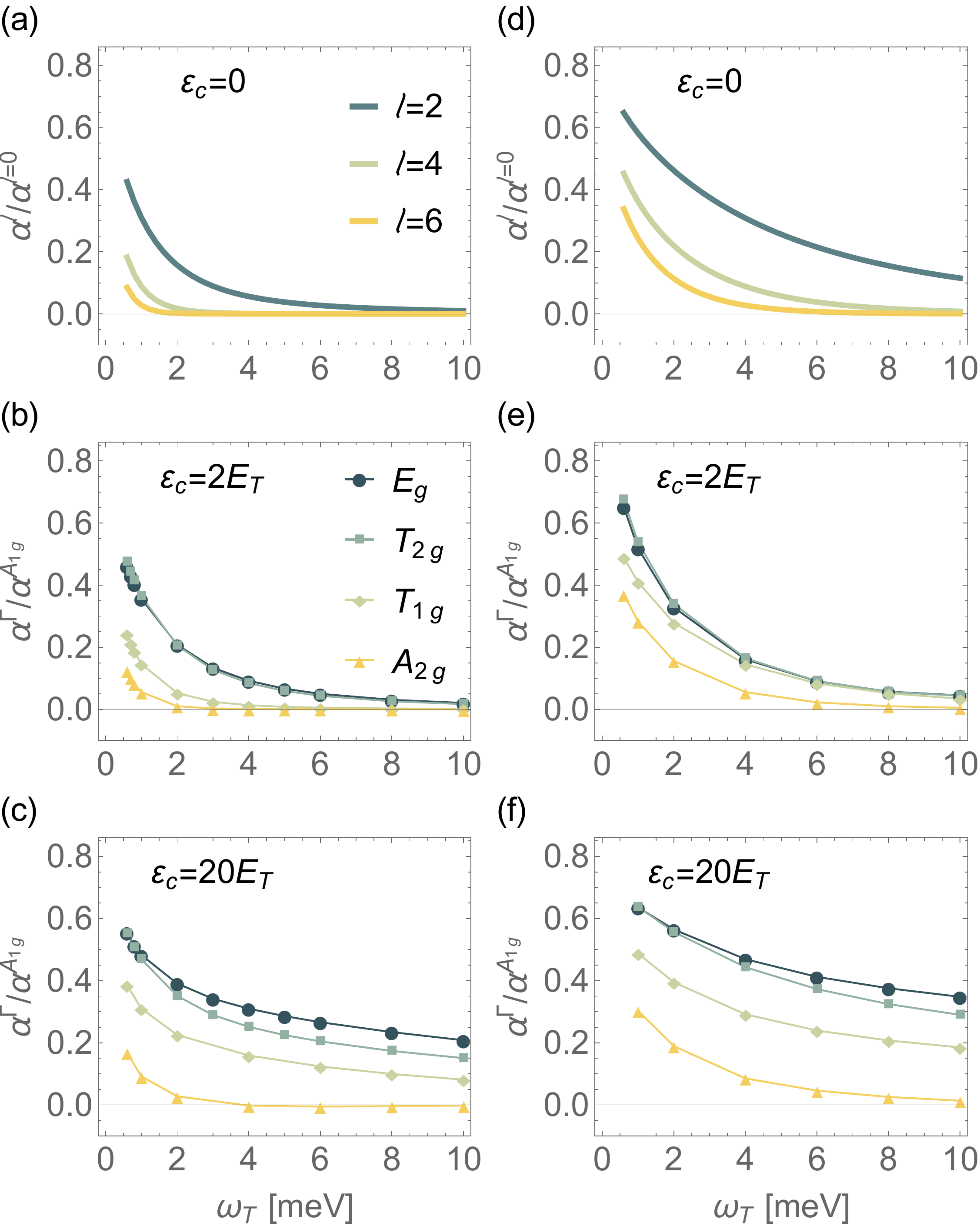}
\caption{Channel competition for various cubic anisotropy values $\varepsilon_{c}$
as indicated in the insets. (a)-(c) $k_{F}=0.05$
and (d)-(e) $k_{F}=0.2$. }
\label{fig:cubic_channels} 
\end{figure}

We now study the pairing instabilities in the other $O_{h}$ even-parity
irreps $\Gamma$ shown in Table \eqref{tab:Ki}. In Fig.~\ref{fig:cubic_channels},
we show the leading eigenvalue $\alpha^{\Gamma}$ corresponding to
each channel (relative to the eigenvalue of the $A_{1g}$ channel)
for different values of the cubic anisotropy $\varepsilon_{c}$ and
two different $k_{F}$ values ($k_{F}=0.05$ for the left panels and
$k_{F}=0.2$ for the right panels). 
Similarly to the $A_{1g}$ case,
we truncate the pairing interaction by considering only the two highest
harmonics shown in Table \eqref{tab:Ki} for each irrep. Panels (a) and (d) recover the results discussed
in Fig. \eqref{fig:isotropic} for the isotropic case, showing that
as $\omega_{T}\rightarrow0$, higher angular momentum instabilities
approach the leading $s$-wave instability. A similar behavior is
seen when the cubic anisotropy is finite, $\varepsilon_{c}\neq0$.
Indeed, in panels (b)-(c) and (e)-(f), the eigenvalues corresponding
to the pairing instabilities in all non-trivial channels become closer
to the eigenvalue of the trivial $A_{1g}$ channel as the TO mode
becomes softer -- although the latter is always larger than the former.
Among the non-trivial irreps, the most favored channels are the $E_{g}$
and $T_{2g}$ ones, usually identified with $d$-wave pairing. We
emphasize however that $E_{g}$ and $T_{2g}$ also have contributions
from $l=4$ angular momentum (``$g$-wave''), as shown in Table
\eqref{tab:Ki}. The reason for the enhancement of the pairing instabilities
in the non-trivial channels seems to be the same for all values of
$\varepsilon_{c}$ (including the isotropic case): as $\omega_{T}\rightarrow0$,
the FE fluctuations become more strongly peaked around $\mathbf{q}=0$,
which tends to favor all pairing states almost equally well.

\section{Discussion and conclusions}

\label{sec:discussion}

In this work, we studied the superconducting instability promoted
by the exchange of FE fluctuations between low-energy fermions in
a cubic system. Focusing on the weak-coupling regime, where the dynamics
of the FE fluctuations is not important, we considered the direct
coupling between the electronic fermionic operators and the FE bosonic
fields mediated by the spin-orbit coupling term \eqref{eq:gamma}~\citep{Fu15}.
In contrast to the dipolar gradient term arising from the electron-phonon
coupling, the main contribution to the pairing interaction comes from
the TO soft mode, as the pairing potential becomes proportional to
$1/\omega_{T}^{2}$. Consequently, we find that $T_{c}$ is enhanced
as the putative FE-QCP is approached, not only in the $s$-wave singlet
channel, but also in all other higher angular momentum even-parity
channels, which become closer competitors to the trivial superconducting
state.

It is important to emphasize that the general problem of weak-coupling
superconductivity caused by the exchange of odd-parity bosonic fluctuations
was generally studied in several recent works \citep{Sau14,Fu15,KoziiFu15,Martin17,Kozii19}.
In all these works, which considered a variety of different spin-orbit
coupling vertices $\Gamma$, it was found that odd-parity and even-parity
channels are close competitors, and sometimes nearly degenerate. Our
work, which considers only the vertex in Eq. \eqref{eq:gamma}, reveals
in addition that as the bosonic mode becomes soft, all orthogonal
even-parity channels become close competitors to the $s$-wave instability.
While here we focused only on the even-parity states, we expect, based
on the results of Ref. \citep{KoziiFu15} for the $l=1$ channel and
for the same vertex $\Gamma$, that different odd-parity channels
will also be enhanced as $\omega_{T}\rightarrow0$. This general phenomenon
of multiple nearby pairing instabilities appearing near a putative
QCP was also observed in the case of superconductivity mediated by
nematic fluctuations \citep{Lederer15,Kang16,Klein18}. Similarly
to those, the FE fluctuations considered here also become strongly
peaked at $q=0$ as the QCP is approached. The key point is that $q=0$
peaked fluctuations, in contrast to the $q$-independent fluctuations
characteristic of the standard electron-phonon interaction, are not
effective in coupling states separated by moderate or large momentum
transfer. As a result, even though the pairing interaction that they
promote is attractive, it does not strongly penalize gap anisotropy.

One of the consequences of the close proximity between the $T_{c}$
values of different even-parity channels as the FE-QCP is approached,
is that certain perturbations may suppress the trivial $s$-wave state
(which always wins in our approach), at the same time that they enhance
non-$s$-wave states -- i.e. states with higher Cooper-pair angular
momentum. For instance, the onsite Coulomb repulsion will certainly
penalize the $s$-wave state but favor nodal states. Whether this
is enough to promote a superconducting transition between two different
pairing states as the FE-QCP is approached remains to be investigated.
We emphasize that our conclusions rely on a weak-coupling calculation
that is valid in a region that excludes the FE-QCP. Although this
excluded region can be small if the coupling constant $g$ is small,
this approximation prevents us from making statements about the nature
of the pairing state at the FE-QCP. At the FE-QCP, the boson dynamics
induced by the coupling to the metal's particle-hole excitations (Landau
damping) becomes crucial.

Our main goal here was to apply this type of spin-orbit-mediated coupling
between odd-parity bosonic fluctuations and electrons to the case
of STO. Experimentally, it is observed that tuning STO towards a putative
FE-QCP via Ca doping, strain, or $^{18}$O substitution leads to an
enhancement of $T_{c}$~\citep{Stucky2016,Rischau2017,Rowley2018,Tomioka2019,Herrera2019,Russell2019}. Theoretically, previous works focused either
on a phenomenological coupling between the FE soft mode and the electrons~\citep{Edge15}
or on the microscopic electron-phonon gradient coupling~\citep{Woelfle18,ArceGamboa18,Kedem18}. While the
non-soft LO mode plays an important role in the latter case, for the
spin-orbit-mediated coupling considered here the pairing interaction
is dominated by the soft TO mode. Our work thus provides an interesting
alternative avenue by which pairing can be enhanced near the FE-QCP
in STO. Of course, there are several features of STO not included
in our analysis, such as the role of dilution and the role of the
multiple bands that cross the Fermi level as doping is changed \citep{Trevisan1,Trevisan2}. Moreover,
the size of the coupling constant $g$ is not known in STO, to the
best of our knowledge. As we said above, even if $g$ is very small,
the interaction can still be large as long as the system is close
enough to the FE-QCP.

An important property of STO that we included in our analysis is the
cubic crystal-field anisotropy of the FE fluctuations, which is not small according
to neutron scattering experiments~\citep{Yamada69}. We find that the main effect of the cubic anisotropy
is to induce an anisotropy in the gap function. Although it transforms
as the trivial $A_{1g}$ representation of the cubic point group $O_{h}$,
the gap consists of an admixture of $s$-wave ($l=0$) and $g$-wave
($l=4$) functions. From this admixture, it follows that the gap displays
maxima or minima at high-symmetry directions $[100]$ and $[111]$.
The gap anisotropy changes non-monotonically as $\omega_{T}$ is suppressed,
changing sign and enhancing in magnitude as the FE-QCP is approached.
Observation of such a gap anisotropy, while challenging, would provide
strong experimental support for the relevance of the mechanism discussed
here to the understanding of the superconducting state of STO. The
best regime to search for such anisotropies would be in the regime
of larger doping concentrations, where the Fermi surface is not too
small.

While our analysis considered a cubic system, STO is actually tetragonal
due to the antiferro-distortive transition it undergoes at about $105$
K. Given that the tetragonal distortion is very small with $c/a=1.00056$~\citep{Lytle1964}, the main results
presented here are unlikely to be changed. One interesting consequence
of such a small tetragonal distortion is that it couples pairing channels
that are otherwise orthogonal in the cubic case. More specifically,
the lattice strain $\varepsilon_{zz}=\partial_{z}u_{z}$, where $\mathbf{u}$
is the lattice displacement, mixes the $A_{1g}$ and the $E_{g}$
states. Because the $E_{g}$ instability becomes a closer competitor
to the $A_{1g}$ instability as the FE-QCP is approached, such a mixing
could lead to an enhancement of $T_{c}$ \citep{Kang14}.

Beyond STO, our work should be relevant for other metallic systems
in which ferroelectric fluctuations are strong and superconductivity
is nearby. A recent work focused on the case of Dirac electrons coupled
to FE fluctuations \citep{Kozii19}. The boson-fermion coupling term
considered in that work is analogous to the one studied here, with
valley degrees of freedom playing the role of spin degrees of freedom.
Interestingly, Ref. \citep{Kozii19} did find a strong enhancement
of $T_{c}$ near the FE-QCP. One could also conceive heterostructures
with substrates that can be continuously tuned across a FE transition,
e.g. $A$TiO$_{3}$ with cation $A$. If a very thin metallic film
is deposited on top of such a substrate, the FE fluctuations of the
latter may provide an additional source of pairing in the metal. Since
the FE fluctuations favor a variety of different pairing channels,
an enhancement of $T_{c}$ would be expected.
\begin{acknowledgments}
We thank A. Balatsky, K. Behnia, A. Chubukov, A. Klein, V. Kozii,
G. Lonzarich, J. Ruhman, and P. Woelfle for insightful discussions.
MNG and RMF were supported by the U. S. Department of Energy through
the University of Minnesota Center for Quantum Materials, under Award
No. DE-SC-0016371. TVT was supported by S\~{a}o Paulo Research Foundation (Fapesp, Brazil) via fellowship 2015/21349-7.
\end{acknowledgments}

\appendix

\section{Perturbative calculation of the interaction due to the cubic anisotropy term $\varepsilon_{c}$}

\label{App:perturbation}

In this section we explicitly derive Eq.~\eqref{eq:Lkpert}, the
expression of the coupling function $L(\hat{k},\hat{k}')$ {[}Eq.~\eqref{eq:Lk}{]}
when treating the cubic crystal field term $\varepsilon_{c}$ perturbatively.
We start with the rotationally invariant case $\varepsilon_{c}=0$
explored in Section~\ref{sec:isotropic}. The phonon dispersions
acquire the simple expressions $\varpi_{a}^{2}(q)=\omega_{a}^{2}+E_{a}^{2}q^{2}$
for the two-fold degenerate transverse mode ($a=T$) and the longitudinal
mode ($a=L$), with polarizations 
\begin{align}
\hat{e}_{T1}(\hat{q}) & =A(\hat{q})\left(-\hat{q}_{z},0,\hat{q}_{x}\right)\label{eq:eT1}\\
 & +B(\hat{q})\left(-\hat{q}_{y}\left|\hat{q}_{x}\right|,(\hat{q}_{x}^{2}+\hat{q}_{z}^{2})\text{sgn}(\hat{q}_{x}),-\hat{q}_{y}\hat{q}_{z}\text{sgn}(\hat{q}_{x})\right)\nonumber \\
\hat{e}_{T2}(\hat{q}) & =C(\hat{q})\left(-\hat{q}_{z},0,\hat{q}_{x}\right)\label{eq:eT2}\\
 & +D(\hat{q})\left(-\hat{q}_{y}\left|\hat{q}_{x}\right|,(\hat{q}_{x}^{2}+\hat{q}_{z}^{2})\text{sgn}(\hat{q}_{x}),-\hat{q}_{y}\hat{q}_{z}\text{sgn}(\hat{q}_{x})\right)\nonumber \\
\hat{e}_{L}(\hat{q}) & =\left(\hat{q}_{x},\hat{q}_{y},\hat{q}_{z}\right)\label{eq:eL}
\end{align}
where the coefficients in the transverse subspace $A(\hat{q})$, $B(\hat{q})$,
$C(\hat{q})$ and $D(\hat{q})$ are chosen to keep the basis $\{\hat{e}_{a}(\hat{q})\}$
orthonormal. The expression of the rotationally invariant coupling
function is then
\begin{equation}
L(x)=\sum_{a=L,T}\frac{E_{T}^{2}}{\omega_{a}^{2}}\frac{1+x}{1+2\eta_{a}^{2}(1+x)}
\end{equation}
which gives the isotropic kernel $\lambda(x)=\frac{\lambda_{0}}{2}\left(L(x)+L(-x)\right)$
in Eq.~\eqref{eq:lambdaa}.

We now introduce the cubic anisotropy term:
\begin{equation}
\varepsilon_{c}^{2}q^{2}\hat{q}_{i}\delta_{ij}\equiv\varepsilon_{c}^{2}q^{2}W_{ij}(\hat{q})
\end{equation}
perturbatively in the bosonic propagator Eq.~\eqref{eq:D0}, and
calculate how the coupling function $L(x)$ is modified through the
changes in the eigenmodes of the propagator. Since in the absence
of the perturbation the transverse subspace {[}Eqs.~\eqref{eq:eT1}-\eqref{eq:eT2}{]}
is doubly degenerate, in order to apply perturbation theory we first
choose the set of coefficients $A(\hat{q})$, $B(\hat{q})$, $C(\hat{q})$
and $D(\hat{q})$ so that the off-diagonal matrix element of the cubic
anisotropy perturbative term vanishes, i.e., $\langle\hat{e}_{T1}(\hat{q})|W|\hat{e}_{T2}(\hat{q})\rangle=0$.
We can now proceed to calculate the modification of the eigen-modes.
First, the perturbation term $W$ lifts the degeneracy of the transverse
modes with modified dispersions $\varpi_{a}^{'2}(\hat{q})\simeq\varpi^2_{a}(q)+\varepsilon_{c}^{2}q^{2}\zeta_{a}(\hat{q})$
up to order $\mathcal{O}(\varepsilon_{c}^{2}q^{2})$ where 
\begin{align}
\zeta_{T1}(\hat{q}) & =\hat{q}_{x}^{2}\hat{q}_{y}^{2}+\hat{q}_{x}^{2}\hat{q}_{z}^{2}+\hat{q}_{y}^{2}\hat{q}_{z}^{2}\label{eq:zetaT1}\\
 & \quad-\sqrt{\hat{q}_{x}^{4}\hat{q}_{y}^{4}+\hat{q}_{x}^{4}\hat{q}_{z}^{4}+\hat{q}_{y}^{4}\hat{q}_{z}^{4}-\hat{q}_{x}^{2}\hat{q}_{y}^{2}\hat{q}_{z}^{2}}\nonumber \\
\zeta_{T2}(\hat{q}) & =\hat{q}_{x}^{2}\hat{q}_{y}^{2}+\hat{q}_{x}^{2}\hat{q}_{z}^{2}+\hat{q}_{y}^{2}\hat{q}_{z}^{2}\label{eq:zetaT2}\\
 & \quad+\sqrt{\hat{q}_{x}^{4}\hat{q}_{y}^{4}+\hat{q}_{x}^{4}\hat{q}_{z}^{4}+\hat{q}_{y}^{4}\hat{q}_{z}^{4}-\hat{q}_{x}^{2}\hat{q}_{y}^{2}\hat{q}_{z}^{2}}\nonumber \\
\zeta_{L}(\hat{q}) & =\hat{q}_{x}^{4}+\hat{q}_{y}^{4}+\hat{q}_{z}^{4}.
\end{align}

The eigenvectors $\hat{e}_{a}(\hat{q})$ in Eqs.~\eqref{eq:eT1}-\eqref{eq:eL}
are also modified by the perturbation term $W$ with $\hat{e}'_{a}(\hat{q})\simeq\hat{e}_{a}(\hat{q})+\varepsilon_{c}^{2}q^{2}u_{a}(\hat{q})$:
\begin{align}
u_{T1}(\hat{q}) & =-\frac{W_{T1,L}(\hat{q})\hat{e}_{L}(\hat{q})}{\varpi_{L}^{2}(q)-\varpi_{T}^{2}(q)}\label{eq:eT1p}\\
u_{T2}(\hat{q}) & =-\frac{W_{T2,L}(\hat{q})\hat{e}_{L}(\hat{q})}{\varpi_{L}^{2}(q)-\varpi_{T}^{2}(q)}\label{eq:eT2p}\\
u_{L}(\hat{q}) & =\frac{W_{T1,L}(\hat{q})\hat{e}_{T1}(\hat{q})+W_{T2,L}(\hat{q})\hat{e}_{T2}(\hat{q})}{\varpi_{L}^{2}(q)-\varpi_{T}^{2}(q)}\label{eq:eLp}
\end{align}

Here, we defined the notation $W_{a,b}(\hat{q})=\langle\hat{e}_{a}|W|\hat{e}_{b}\rangle$
for the perturbation matrix elements. We can thus calculate the correction
of the numerator in the coupling function Eq.~\eqref{eq:L_final},
$\Upsilon'_{a}(\hat{q})\simeq\Upsilon_{a}(q)+\varepsilon_{c}^{2}q^{2}\Phi_{a}(\hat{q})$
with
\begin{align}
\Phi_{T1}(\hat{k},\hat{k}\,') & =-\frac{2W_{T1,L}(\hat{q})}{\varpi_{L}^{2}(q)-\varpi_{T}^{2}(q)}\times\\\nonumber
&\quad\sum_{i,j=x,y,z}u_{L,i}(\hat{q})u_{T1,j}(\hat{q})(\hat{k}_{i}+\hat{k}'_{i})(\hat{k}_{j}+\hat{k}'_{j})\\
\Phi_{T2}(\hat{k},\hat{k}\,') & =-\frac{2W_{T2,L}(\hat{q})}{\varpi_{L}^{2}(q)-\varpi_{T}^{2}(q)}\times\\\nonumber
&\quad\sum_{i,j=x,y,z}u_{L,i}(\hat{q})u_{T2,j}(\hat{q})(\hat{k}_{i}+\hat{k}'_{i})(\hat{k}_{j}+\hat{k}'_{j})\\
\Phi_{L}(\hat{k},\hat{k}\,') & =-\left[\Phi_{T_{1}}(\hat{k},\hat{k}\,')+\Phi_{T_{2}}(\hat{k},\hat{k}\,')\right],
\end{align}
and $\hat{q}=(\hat{k}-\hat{k}\,')/\sqrt{2(1-\hat{k}\cdot\hat{k}\,')}$. Because of the form of the
longitudinal polarization $\hat{e}_{L}(\hat{q})$ {[}Eq.~\eqref{eq:eL}{]},
the following equality is fulfilled,
\begin{widetext}
\begin{equation}
\sum_{i=x,y,z}u_{L,i}(\hat{q})u_{a,i}(\hat{q})(\hat{k}_{i}+\hat{k}'_{i})^{2}=-\sum_{i\neq j}u_{L,i}(\hat{q})u_{a,j}(\hat{q})(\hat{k}_{i}+\hat{k}'_{i})(\hat{k}_{j}+\hat{k}'_{j})\longrightarrow\Phi_{a}(\hat{q})=0.
\end{equation}
\end{widetext}

Therefore, the leading order correction of $\Upsilon_{a}(\hat{q})$
vanishes and, as mentioned in the main text, only the correction to
the eigenvalues $\varpi_{a}'(\hat{q})$ contributes in this order
to the modified expression of the pairing interaction, given by Eq.~\eqref{eq:Lkpert}.

\section{Non-zero $c_{04}^{A_{1g}}$ mixing term}

\label{App:c40}

We focus on the simplest limit of the lengthy expressions Eqs.~\eqref{eq:lambdacT}-\eqref{eq:lambdacL}:
$\omega_{a}^{2}\ll E_{a}^{2}k_{F}^{2}$, so we can take $\eta_{a}^{2}\approx0$
in the denominators of both expressions (note that for the massive
longitudinal mode this is a very good approach, but for the soft transverse
mode it is just one of the possible limits that depend on the ratio
$\frac{\omega_{T}}{E_{T}k_{F}}$). For simplicity, we also set $x=0$,
which we can do without loss of generality in order to show that $c_{40}^{A_{1g}}\neq0$.
The sign of $c_{40}$, however, depends on the contribution of all
$x$. Under these approximations the expressions become: 
\begin{widetext}
\begin{align}
\lambda_{c,T}(\hat{k},\hat{k}') & \simeq-\lambda_{0}\left(\frac{E_{T}}{\omega_{T}}\right)^{2}\left(\frac{\varepsilon_{c}k_{F}}{\omega_{T}}\right)^{2}\left[\hat{k}_{x}^{2}\hat{k}_{y}^{2}+\hat{k}_{x}^{2}\hat{k}_{z}^{2}+\hat{k}_{y}^{2}\hat{k}_{z}^{2}+\hat{k}_{x}^{'2}\hat{k}_{y}^{'2}+\hat{k}_{x}^{'2}\hat{k}_{z}^{'2}+\hat{k}_{y}^{'2}\hat{k}_{z}^{'2}+1-3\left(\hat{k}_{x}^{2}\hat{k}_{x}^{'2}+\hat{k}_{y}^{2}\hat{k}_{y}^{'2}+\hat{k}_{z}^{2}\hat{k}_{z}^{'2}\right)\right]\label{eq:lT}\\
\lambda_{c,L}(\hat{k},\hat{k}') & \simeq-\frac{\lambda_{0}}{2}\left(\frac{E_{T}}{\omega_{L}}\right)^{2}\left(\frac{\varepsilon_{c}k_{F}}{\omega_{L}}\right)^{2}\left[\hat{k}_{x}^4+\hat{k}_{y}^4+\hat{k}_{z}^4+\hat{k}_{x}^{'4}+\hat{k}_{y}^{'4}+\hat{k}_{z}^{'4}+6\left(\hat{k}_{x}^2\hat{k}_{x}^{'2}+\hat{k}_{y}^2\hat{k}_{y}^{'2}+\hat{k}_{z}^2\hat{k}_{z}^{'2}\right)\right]\text{ .}\label{eq:lL}
\end{align}
\end{widetext}

Comparing Eqs.(\ref{eq:lT}) and (\ref{eq:lL}) with Eq.~\eqref{eq:K4}, we immediately see their similarities
with the cubic harmonics. More concretely we can rewrite the expressions
as
\begin{widetext}

\begin{align}
\lambda_{c,T}(\hat{k},\hat{k}') & \propto-\frac{1}{2}\sum_{i>j}\left[(\hat{k}_{i}-\hat{k}'_{i})^{2}(\hat{k}_{j}-\hat{k}'_{j})^{2}+(\hat{k}_{i}+\hat{k}'_{i})^{2}(\hat{k}_{j}+\hat{k}'_{j})^{2}\right]\approx\frac{8\pi}{5}\begin{pmatrix}K_{0}^{A_{1g}} & K_{4}^{A_{1g}}(\hat{k})\end{pmatrix}\begin{pmatrix}-\frac{8}{3} & \frac{1}{\sqrt{21}}\\
\frac{1}{\sqrt{21}} & 0
\end{pmatrix}\begin{pmatrix}K_{0}^{A_{1g}}\\
K_{4}^{A_{1g}}(\hat{k}')
\end{pmatrix}\\
\lambda_{c,L}(\hat{k},\hat{k}') & \propto-\frac{1}{4}\sum_{i}\left[(\hat{k}_{i}-\hat{k}'_{i})^{4}+(\hat{k}_{i}+\hat{k}'_{i})^{4}\right]\approx\frac{8\pi}{5}\begin{pmatrix}K_{0}^{A_{1g}} & K_{4}^{A_{1g}}(\hat{k})\end{pmatrix}\begin{pmatrix}-4 & -\frac{1}{\sqrt{21}}\\
-\frac{1}{\sqrt{21}} & 0
\end{pmatrix}\begin{pmatrix}K_{0}^{A_{1g}}\\
K_{4}^{A_{1g}}(\hat{k}')
\end{pmatrix} \text{ ,} 
\end{align}
\end{widetext}

\noindent where we truncated the expansions in cubic harmonics to leading order. Note that in both cases, the angular momentum mixing coefficient $c_{40}^{A_{1g}}\neq0$.

\bibliography{references}

\end{document}